\documentstyle [12pt] {article}
\catcode`\@11
\def\section{
    \setcounter{equation}{0}
\@startsection {section}{1}{\z@}{-3.5ex plus -1ex minus -.2ex}
{2.3ex plus .2ex}{\large\bf}
}
\catcode`\@13
\newcommand{\nummer}[1]{\hskip 12 true cm #1 \par}
\newcommand{\netnum}[1]{\vspace{-14pt}\hskip 12 true cm #1 \par}
\newcommand{\monat}[1]{\hskip 12 true cm #1
                       \par \vspace*{1 cm}}
\newcommand{\titel}[1]{{\renewcommand{\thefootnote}{\fnsymbol{footnote}}
                       \Large\bf\vskip 0 true cm
                       \begin{center}#1\end{center}
                       \setcounter{footnote}{0}}
                       \normalsize\vskip 1.2 true cm}
\newcommand{\autor}[1]{{
                       \renewcommand{\thefootnote}{\arabic{footnote}}
                       \begin{center} {\large #1 }\end{center}}
                       \setcounter{footnote}{0}}
\newcommand{\adresse}[1]{\vspace*{-1.1 true cm}\begin{center} {\it #1 }
                         \end{center}
                         \vskip 0.5cm}

\newcommand{\bye}{\end{document}}
\newcommand{\be}{\begin{equation}}
\newcommand{\ee}{\end{equation}}
\newcommand{\bes}{\begin{eqnarray}}
\newcommand{\ees}{\end{eqnarray}}
\newcommand{\ema}{\end {array} \right)}
\newcommand{\nin}{\kern 0.1 em \in\kern -0.80em /}
\newcommand{\pslash}{\kern 0.1 em p\kern -0.45em /}
\newcommand{\dslash}{\kern 0.1 em \partial\kern -0.55em /}
\newcommand{\sla}[1]{\kern 0.1 em #1\kern -0.55em /}
\newcommand{\ra}{\rightarrow}

\newcommand{\lra}{\longrightarrow}

\newcommand{\R}{{I\kern -0.22em R\kern 0.30em}}
\newcommand{\N}{{I\kern -0.22em N\kern 0.30em}}
\newcommand{\C}{\mbox{\kern 0.20em \raisebox{0.09ex}{\rule{0.08ex}{1.22ex}}
                \kern -0.60em C\kern 0.30em}}
\newcommand{\Z}{{\sf Z\kern -0.40em Z\kern 0.30em}}
\newtheorem{theo}{Theorem}

\newtheorem{lemm}{Lemma}
\newcommand{\qed}{$\; \Box$}
\begin{document}
\newcommand{\OD}{\Omega_D }
\newcommand{\cA}{\cal A}
\newcommand{\cAs}{\cal A_s }
\newcommand{\vA}{\mbox{{\bf A}} }
\newcommand{\vB}{\mbox{{\bf B}} }
\newcommand{\bc}{\mbox{{\bf c}}}
\newcommand{\vz}{\mbox{{\bf z}}}
\newcommand{\cC}{\cal C}
\newcommand{\cE}{\cal E }
\newcommand{\cEs}{\cal E_s }
\newcommand{\vE}{\mbox{{\bf E}} }
\newcommand{\cEt}{\tilde{\cal E} }
\newcommand{\cEts}{\tilde{\cal E_s} }
\newcommand{\cF}{\cal F }
\newcommand{\vF}{\mbox{{\bf F}} }
\newcommand{\vG}{\mbox{{\bf G}} }
\newcommand{\cH}{\cal H }
\newcommand{\cHs}{\cal H_s }
\newcommand{\cHp}{\cal H_\pi }
\newcommand{\cHps}{{\cal H_s}_\pi }
\newcommand{\cJ}{\cal J }
\newcommand{\cK}{\cal K }
\newcommand{\cL}{\cal L }
\newcommand{\cM}{\cal M }
\newcommand{\cN}{\cal N }
\newcommand{\cP}{\cal P }
\newcommand{\cB}{{\cal B}(\cH) }
\newcommand{\cBs}{{\cal B}(\cHs) }
\newcommand{\bo}[1]{\mbox{\boldmath$#1$}}
\newcommand{\OA}{\Omega \cA }
\newcommand{\ON}{\Omega_{\cN}}
\newcommand{\pds}{\pi_{D_s}}
\newcommand{\Hom}{\mbox{Hom}_{\cA}(\cE,\cE\otimes_{\cA}\OD\cA)}
\newcommand{\Homa}{\mbox{Hom}_{\cA}(\cE,\cE\otimes_{\cA}\OD^1\cA)}
\newcommand{\Homb}{\mbox{Hom}_{\cA}(\cE,\cE\otimes_{\cA}\OD^2\cA)}
\newcommand{\Homk}{\mbox{Hom}_{\cA}(\cE,\cE\otimes_{\cA}\OD^k\cA)}
\newcommand{\Homs}{\mbox{Hom}_{\cA}(\cE,\cE\otimes_{\cA}\ON^1\cA)}
\newcommand{\Homt}{\mbox{Hom}_{\cA}(\cE,\cE\otimes_{\cA}\cA dt)}
\newcommand{\Homss}{\mbox{Hom}_{\cA}(\cE,\cE\otimes_{\cA}\ON^2\cA)}
\newcommand{\Homst}{\mbox{Hom}_{\cA}(\cE,\cE\otimes_{\cA}\ON^1\cA dt)}
\newcommand{\Homks}{\mbox{Hom}_{\cA}(\cE,\cE\otimes_{\cA}\ON^k\cA)}
\newcommand{\Homkt}{\mbox{Hom}_{\cA}(\cE,\cE\otimes_{\cA}\ON^{k-1}\cA dt)}
\newcommand{\diva}[1]{\mbox{{\bf #1}} }

\begin{titlepage}
\nummer{MZ-TH/94-26}
\netnum{gr-qc/9409193}
\monat{\today}
\titel{Hamilton Formalism in Non-Commutative Geometry\footnote{
{\rm Work supported in part by the PROCOPE agreement between the
University of Aix-Marseille and the Johannes Gutenberg-Universit\"at of Mainz.
}}}
\autor{Wolfgang Kalau\footnote{
e-mail: kalau{\char'100}dipmza.physik.uni-mainz.de}}
\adresse{Johannes Gutenberg Universit\"at\\
Institut f\"ur Physik\\
55099 Mainz}
\begin{abstract}
We study the Hamilton formalism for Connes-Lott models, i.e., for
Yang-Mills theory in non-commutative geometry. The starting point is an
associative $*$-algebra $\cA$ which is of the form $\cA=C(I,\cAs)$ where
$\cAs$ is itself a associative $*$-algebra. With an appropriate choice of a
k-cycle over $\cA$ it is possible to identify the time-like part of the
generalized differential algebra constructed out of $\cA$. We define the
non-commutative analogue of integration on space-like surfaces via the
Dixmier trace restricted to the representation of the space-like part
$\cAs$ of the algebra. Due to this restriction it possible to define the
Lagrange function resp.~Hamilton function also for Minkowskian space-time.
We identify the phase-space and give a definition of the Poisson bracket
for Yang-Mills theory in non-commutative geometry. This general formalism
is applied to a model on a two-point space and to a model on Minkowski
space-time $\times$ two-point space.
\end{abstract}
\end{titlepage}
\section{Introduction}
In the last few years it has turned out that A.~Connes' non-commutative
geometry provides a framework which allows for new qualitative insights in the
spontaneous symmetry breaking mechanism of Yang Mills theories. The
cornerstone of this approach is a generalization of the algebra of
differential forms and its corresponding differential. This has been used to
construct models for the electroweak interaction \cite{colo,cobuch} and
Grand Unification \cite{chams1,chams2}. Since the generalization of
the differential algebra and its differential is not unique there are
alternative models for the electroweak interaction, like the one developed by
the Marseille and Mainz groups \cite{cev,hps1,hps2}. However, all models have
in common that the Higgs field is interpreted as a part of the generalized
connection form, although the precise form of the Higgs potential depends on
the model chosen.

Another feature, which is common to all models so far, is that they are purely
classical models, i.e. non-commutative geometry has been used to derive
classical actions. In this approach some coupling constants, like
the Higgs mass and the top mass in the
Connes-Lott model, appear naturally restricted. However, such relations at the
classical level cannot be translated to relations at the quantum level in an
obvoius way \cite{AGBM}. The reason for this is that it is not known so far
how to quantize a theory in the framework of non-commutative geometry and for
the usual quantization procdedure it does not matter if some coupling constants
of the classical action are fixed by hand or by some general principles of
non-commutative geometry. Therefore it seems desirable to have a translation
of the usual quantization procedure into the language of non-commutative
geometry in order to get new insight in quantized Yang-Mills theory.

The generalization of geometry to non-commutative geometry is achieved by
translating geometrical concepts into an algebraic language where
conventional geometry corresponds to commutative algebras. The generalization
is then obtained by extending those concepts to non-commutative algebras.

The quantization procedure which is closely related to algebra is the
canonical quantization method. This approach to quantum theory is based on
the Hamilton formalism. The purpose of this article is to develope an Hamilton
formalism for (generalized) Yang-Mills theories in non-commutative geometry
as they were introduced in \cite{colo,cobuch}.

This article is organized as follows. In sect.~2 we give a motivation for the
structure\linebreak
 $\cA=C(I,\cAs)$ of the associative $*$-algebra $\cA$ which is the
starting point for the derivation of Yang-Mills theory in non-commutative
geometry. The universal differential enveloping algebra and the concept of
finitely summable k-cycles are briefly reviewed in sect.~3 where we construct
a k-cycle which is appropriate for our purpose. In sect.~4 the generalized
differential algebra $\OD\cA$ of A.~Connes\cite{cobuch} is discussed
where we use the structure on $\cA$ and the k-cycle, introduced in the previous
sections, to show that there is a split
of $\OD\cA$ into a ``space-like'' and a ``time-like'' part. The trace theorem
of A.~Connes \cite{co} is used in sect.~5 to define an inner product on
$\OD\cA$. This definition differs from the usual definition in the sense that
it corresponds to an integration on a ``space-like'' surface. As a consequence
it is possible to define it also on space-time geometries with Minkowski
metric. After a brief review of Yang-Mills theory in non-commutative geometry
as it was introduced by A.~Connes and J.~Lott \cite{colo,cobuch}, the
Lagrange function and the Hamiltonian for Yang-Mills theory are defined in
sect.~6. The formal construction ends with the definition of the Poisson
bracket and time evolution in sect.~7. In sect.~8 the formalism is applied
to two examples, namely to a discrete space  and to Yang-Mills theory
with symmetry breaking. The article ends with some conclusions in
sect.~9.
\section{The Algebra $\cA$}
Hamilton formalism is related to Cauchy surfaces in space-time and
the separation of time which implies that the space-time manifold
$M$ has the topology
\be
M = \R \times \Sigma
\ee
where $\R$ corresponds to time and $\Sigma$ to a (compact) space-like
manifold. As consequence the corresponding $C^*$-algebra of continous
functions (vanishing at infinity) $C_0(M)$ is of the form
\be
C_0(M) = C_0(\R) \otimes C(\Sigma) = C_0(\R, C(\Sigma)) \label{prod}
\ee
where $C_0(\R) \otimes C(\Sigma)$ denotes the completition of the
algebraic product of $C(\R)$ and $C(\Sigma)$ and $C_0(\R, C(\Sigma))$,
or more generally $C_0(\R , \cA)$, is the algebra of continous functions
over $\R$ with values in $C(\Sigma)$ resp.~with values in some normed
algebra $\cA$.

The starting point of A.~Connes' generalization of differential forms is
an associative $*$-algebra $\cA$ (a subalgebra of a $C^*$-algebra).
Equation (\ref{prod}) motivates us to
require that $\cA$ has some additional structure which allows to
introduce ``time'' to the formalism of generalized differential forms.
Thus we postulate that
\be
\cA = C(I,\cAs)\;\; ,\label{gprod}
\ee
where $I$ is either $\R$ or $S^1$ and $\cAs$ is a normed associative
$*$-algebra with unit, possessing a finitely summable k-cycle.
If $\cAs$ is a $C^*$ algebra we have
\be
\cA = C(I,\cAs) = C(I)\otimes\cAs
\ee
where $C(I)\otimes\cAs$ again denotes the completition of the algebraic
product of $C(I)$ and $\cAs$. Since $\cAs$ has a unit we can identify $C(I)$,
the algebra of continous functions on $I$, as a subalgebra of $\cA$ by
\be
\begin{array}{rclr}
i_t   & : & C(I) \lra \cA & \\
    &   &               & \\
i_t(f)& = & f\otimes 1_s  & f\in C(I)
\end{array}
\ee
where $1_s$ denotes the unit element in $\cAs$.

We shall assume that $\cA$ has a unit element. If $I$ is compact
(i.e. $I=S^1$) then $C(I)$ and therefore also $\cA$ has a unit element.
If $I=\R$ then $C_0(I)$ does not have a unit. However we can always
formally add a unit element to $C_0(I)$ which induces a unit element in $\cA$.
Furhtermore we can use the unit element $1_t$ of $C(I)$ to identify $\cAs$
as a subalgebra of $\cA$:
\be
\begin{array}{rclr}
i_s   & : & \cAs \lra \cA & \\
    &   &               & \\
i_s(a)& = & 1_t\otimes a  & f\in \cAs\;\; .
\end{array}
\ee

\section{The Universal Differential Envelope and the k-cycle over $\cA$}
In this and the subsequent section we follow A.~Connes construction of
generalized differential forms \cite{cobuch}. However, we will focus on the
structure of $\cA= C(I,\cAs)$ which will lead to a
``time-split'' in the generalized differential algebra. For details of the
general construction we refer to \cite{cobuch,Kbuch,GBV}.

The first step is to construct a bigger algebra
$\Omega\cA$ by associating to each element $A\in\cA$ a symbol
$\delta A$. $\Omega\cA$ is the free algebra generated by the symbols $A$,
$\delta A$, $A\in\cA$ modulo the relation
\be
\delta (AB)=\delta A\, B + A\delta B\;\; .\label{gl-4}
\ee
With the definition
\be
\begin{array}{rcl}
\delta(A_0\delta A_1\cdots\delta A_k) & \;:= &
\;\delta A_0\,\delta A_1\cdots\delta A_k \\
 & & \\
\delta(\delta A_1\cdots\delta A_k) & \;:= & 0
\end{array}
\ee
$\Omega\cA$ becomes a $\N$-graded differential algebra with the odd
differential $\delta$, $\delta^2=0$. $\OA$ is called the universal
differential envelope of \cA.
By defining
\be
{\delta(A)}^*=-\delta(A^*)
\ee
the $*$-operation is extended uniquely to $\OA$.

The next element in the construction is the k-cycle $(\cH, D)$ over $\cA$.
It consists of a Hilbert space $\cH$ with a faithful $*$-representation
$\pi$
\be
\pi : \cA \lra \cB
\ee
where $\cB$ denotes the algebra of bounded operators on $\cH$. The second
part of the k-cycle is an unbounded self-adfoint operator $D$ on $\cH$. Since
the k-cycle should also reflect the structure given by eq.(\ref{gprod}) let us
first discuss the representation $\pi$ a little bit further before we come
to structure of $D$. However, the main strategy will be to construct the
k-cycle $(\cH, D)$ over $\cA$ out of k-cycles $(\cHs, D_s)$ over $\cAs$.

Suppose $\cHs$ is a (seperable) Hilbert space with an inner product
${(\cdot ,\cdot)}_s$ and a faithful $*$-representation
${\tilde{\pi}}_s$
\be
{\tilde{\pi}}_s : \cAs \lra \cBs\;\; .
\ee
$\cH_s$ can be extended to a bigger Hilbert space
\be
\cH = L_2(I,\cHs)
\ee
with the inner product
\be
(\Psi,\Phi) = \int_I \! dt {(\Psi(t),\Phi(t))}_s\;\; .
\ee
The representation ${\tilde{\pi}}_s$ on $\cHs$ induces a representation $\pi_s$
on $\cH$ of $\cAs$
\be
\pi_s : \cAs \lra C(I,\cBs) \subset \cB
\ee
by indentifying $\cBs$ with the subalgebra of operators in $C(I,\cBs$ which
are constant in $t\in I$. There is also a representation $\pi_t$ of $C(I)$
\be
\begin{array}{rcl}
\pi_t   & : & C(I) \lra C(I,\cBs) \\
    &   &                         \\
\pi_t(f)& = & f\, {id}_s \;\; ,\;\; f\in C(I)
\end{array}
\ee
where ${id}_s$ denotes the unit element in $\cBs$. Because of eq.(\ref{gprod})
these two representations induce a faithful $*$-representation $\pi$ of $\cA$
\be
\pi : \cA \lra C(I,\cBs)
\ee
with
\be
\pi(f\otimes a)=\pi_t(f)\pi_s(a)=\pi_s(a)\pi_t(f)\;\; ,\;\; f\otimes a\in \cA
\ee
Strictly speaking, $\pi_s$ and $\pi_t$ define a representation of a dense
subalgebra of $\cA$, which can be extended to a representation of $\cA$.

Let us now turn to the second element of the k-cycle, the operator $D$
on $\cH$. The general conditions to be fulfilled by this operator are
\cite{cobuch}
\begin{itemize}
\item[{\bf i.}] $D$ is self-adjoint;
\item[{\bf ii.}] $[D,\pi(A)]$ is a bounded operator;
\item[{\bf iii.}] $D$ is unbounded with a compact inverse (modulo finite rank
operators) such that $|D|^{-1}$ is
 $d^+$ summable for some $d\in\N$;
\end{itemize}
If $\cA$ is a $C^*$ algebra condition {\bf ii.} holds only on a dense
subalgebra of $\cA$ in general. Therefore we denote in the following by $\cA$
a dense subalgebra of a $C^*$ algebra such that {\bf iii.} holds for any
element of $\cA$, i.e.~$\cA=C^\infty(I,\cAs)$, where $\cAs$ is also a
suitable dense subalgebra of a $C^*$ algebra.
However, since $D$ is closely related to the metric structure of the
underlying manifold, which is also the case for
non-commutative geometries\cite{cobuch}\footnote{In fact, if $D$ is a Dirac
operator it is possible to construct a gravity-action by taking the Wodzicki
residue of an appropriate inverse power of $D$\cite{coLS,kagr,KW}.},
we have to impose further conditions on $D$. They should reflect the topology
which is encoded in the structure (\ref{gprod}) of $\cA$. This motivates the
additional requirement that $D$ is the sum of two operators
\be
D= D_t + D_s
\ee
with
\begin{itemize}
\item[{\bf iv.}]
$
[D_t,\pi_s(a)]=0\;\;\; \mbox{and} \;\;\; [D_s, \pi_t(f)]=0\;\; ,\;\;
\forall f\in C(I)\;\; ,\;\;  \forall a\in \cAs\;\; ;
$
\item[{\bf v.}]
$
[D_t,\pi_t(f)][D_s,\pi_s(a)]+[D_s,\pi_s(a)][D_t,\pi_t(f)]=0\;\; ,\;\;
\forall f\in C(I)\;\; ,\;\; \forall a\in \cAs\;\; ;$
\item[{\bf vi.}] $D_s \in C^\infty(I, {\cal O}_s)$\footnote{Note, this implies
the second part of {\bf iv.}.}, (where ${\cal O}_s$ denotes the algebra of
operators on $\cHs$) thus $D_s$ is as a smooth 1 parameter
family of operators on $\cHs$. $D_s(t)$ fulfills conditions {\bf i.}-{\bf iii.}
with $\cA$ replaced by $\cAs$, $\cH$ replaced by $\cHs$. In other words
$(\cHs, D_s(t))$ is a smooth 1 parameter family of k-cycles over $\cAs$.
\end{itemize}
We now show how a k-cycle $(\cH, D)$ over $\cA$ can be constructed out of a
1-parameter family of k-cycles $(\tilde{\cHs}, \tilde{D_s}(t))$, $t\in I$, over
$\cAs$. Having the $3+1$-dimensional case in mind, we do not assume that there
is a grading on $\cHs$, i.e.~an automorphism $\gamma$ with $\gamma^2=1$ and
$[\gamma,\tilde{D_s}]_+ =0$\footnote{$[ \cdot, \cdot]_+$ denotes the
anti-commutator} and $[\gamma, \pi_s] = 0$. However, a substitute for this
automorphism can always be constructed by extending $\tilde{\cHs}$, which also
allows to drop the condition that $\tilde{D}_s$ is self-adjoint. We extend
the Hilbert space by $\C^{\!2}$:
\be
\begin{array}{rcl}
\cHs & = & \tilde{\cHs}\otimes \C^{\!2}\\
    &   &                         \\
\pi_s&= & \tilde{\pi_s}\otimes 1_{\C^{\!2}}
\end{array}\label{c2-ext}
\ee
and
\be
D_s =
\left( \begin{array}{cc}
0 & \tilde{D_s} \\
  &             \\
\tilde{D_s}^\dagger & 0
\end{array} \right) \;\; .\label{dsdef}
\ee
The Hilbert space $\cHs$ and the representation $\pi_s$ can be extended
to a Hilbert space $\cH$ and a representation $\pi$ of $\cA$ in the above
mentioned manner. What is still missing is the operator $D$. More precisely
the part $D_t$ of $D$ has to be specified. A natural choice is
$D_t\sim \partial_t$. However, condition {\bf v.} has to be
taken into account. Therefore we introduce an element $\gamma^0 \in \cB$
(a substitute for the grading) with the following properties:
\be
\gamma^0=\left(\begin{array}{cc}
              \tilde{\gamma} &     0        \\
                           &              \\
                    0      & -\tilde{\gamma}
\end{array}\right)\;\; ; \label{blockdia}
\ee
\be
{\gamma^0}^2 = \pi(N)\; , \; N\in\cA \;\; ;\;\;
{[\gamma^0, \pi(A)]} =0 \; , \;\forall A\in \cA \;\; ;\;\;
{\gamma^0}^{-1}\;\; \hbox{exists}\label{etacon}
\ee
and
\be
{\gamma^0}^\dagger = -\gamma^0\;\; ,\label{eukl}
\ee
where the same block structure as in eq.(\ref{c2-ext}) is used.
Such an element always extists since $\cA$ has an unit element.
We now define $D_t$ by
\be
D_t = \gamma^0\partial_t \;\; .\label{dtdef}
\ee
The anti self-adjointness of $\gamma^0$ ( eq.(\ref{eukl})) ensures the
self-adjointness of $D=D_t+D_s$.
It is straightforward to check that
for this choice of $D=D_t+D_s$, with $D_t$ resp.~$D_s$ defined as in
eq.(\ref{dtdef}) resp.~(\ref{dsdef}), and $\cH$ $\; (\cH, D)$ is a k-cycle
over $\cA=C^\infty(I,\cAs)$ which fulfills conditions {\bf ii.} and
{\bf iv.}-{\bf vi.}. Condition {\bf iii.} is not needed for the definition
of the generalized differential algebra. It is crucial for the definition
of the operator theoretic substitute for integration. However, since we are
only interested in a substitute for integration on ``space-like'' surfaces
we can replace this condition on $D$ by an analogous condition on $D_s$
({\bf vi.}). However, this condition on $D$ and the self-adjointness of $D$
is related to the Euklidean signature of the metric of the underlying
manifold. As we shall see in sect.\ref{mink}, with the choice
$\gamma^0={\gamma^0}^\dagger$ one obtains an operator $D$ corresponding to an
underlying manifold with Minkowski metric.

\section{The Generalized Differential Algebra}
Having introduced the generalized differential algebra $\OA$ of $\cA$
and a k-cycle $(\cH,D)$ over $\cA$ we can now put these elements together
in order to define a generalized differential algebra as it was introduced
by A.~Connes \cite{cobuch}.
We begin with extending the $*$-representation $\pi$ to a $*$-representation
of the algebra $\OA$
\be
\begin{array}{rcl}
\pi_D & :  & \OD \lra \cB \\
      &    &          \\
\pi_D (A_0\delta A_1 \cdots \delta A_k) & = &
\pi(A_0)[D,\pi(A_1)]\cdots [D,\pi(A_k)]\;\; .
\end{array}\label{pd}
\ee
However, there is another possibility to extend $\pi$ to a representation
of $\OA$ which is useful for our purpose:
\be
\begin{array}{rcl}
\pds & :  & \OD \lra \cB \\
      &    &          \\
\pds(A_0\delta A_1 \cdots \delta A_k) & = &
\pi(A_0)[D_s,\pi(A_1)]\cdots [D_s,\pi(A_k)]\;\; .
\end{array}\label{pds}
\ee
Obviously the kernel of $\pds$ is much bigger than the kernel of $\pi_D$.
For instance $\delta C^\infty(I) \subset \Omega^1\cA$ is contained in the
kernel of $\pds$ because only the ``space-like'' part $D_s$ of $D$ is used
in the definition eq.(\ref{pds}) of $\pds$.

On the images of $\pi_D$ resp.~$\pds$ the differential $\delta$ on $\OA$
does not induce well defined differentials. Therefore one has to devide
out two sided graded differential ideals. For $\pi_D$ such an ideal
is given by
\be
\begin{array}{rcl}
\cJ^k & = & \ker\pi_D \cap \Omega^k\cA + \delta(\ker\pi_D \cap
\Omega^{k-1}\cA)\\
        &   & \\
\cJ   & = & \bigoplus^\infty_{k=1} \cJ^k\;\; . \label{jdef}
\end{array}
\ee
On the quotient algebra $\OD\cA$, which is defined as
\be
\begin{array}{rcl}
\OD^k\cA & = & {\Omega^k\cA\over \cJ^k}\\
         &   &                           \\
\OD\cA   & = & \bigoplus_{k=1}^\infty \OD^k\cA
\end{array}
\ee
there is a differential $d$ with $d^2=0$ which is uniquely defined by the
differential $\delta$ on $\OA$ as
\be
d(\sigma_D\pi_D(\omega))=
\sigma_D\pi_D(\delta\omega)\;\;,\;\omega\in\OD
\ee
where $\sigma_D$ denotes the map
\be
\sigma_D:\pi_D(\Omega^k\cA)\lra\OD^k\cA\;\; .
\ee
Thus $\OD\cA$ is a generalized graded differential algebra \cite{cobuch}.

Of course, it is now possible to define the differential ideal associated to
$\pds$ in a completely analogous way as for $\pi_D$. This would also lead to a
differential algebra with a differential which is uniquely defined by the
differential on $\OA$. However, such a differential algebra would not have an
interpretation as the ``space-like'' part of $\OD\cA$ in general. Therefore one
has to divide out a bigger differential ideal. One is led to the correct ideal
by the two lemmas following the next little preparing lemma.
\begin{lemm}\label{le0}
For $I= S^1$ there is an
\be
\eta= \sum_i f^{(i)}\delta g^{(i)}\in\Omega^1\cA\; , f^{(i)},g^{(i)}\in C(I)
\label{etas1}
\ee
such that
\be
\pi_D(\eta)=\gamma^0\;\; .\label{etas1a}
\ee
If $I=\R$  there is a sequence
\be
\eta_n=\sum_if^{(i)}_n\delta g^{(i)}_n\in\Omega^1\cA\; ,
f^{(i)}_n,g^{(i)}_n\in C(i)
\ee
such that
\be
\lim_{n\ra\infty} \pi_D(\eta_n) = \gamma^0
\ee
in the strong operator topology of $\cB$.
\end{lemm}
{\bf Proof:}
If $I=S^1$ let $U_i$ be a finite open cover of $S^1$, $f_i$ the corresponding
partition of unity and $g_i$ some smooth functions on $S^1$ with
$\partial_tg_i=1\; ,\forall t\in U_i$ then $\eta$ defined as in
eq.(\ref{etas1})
fulfills eq.(\ref{etas1a}).

If $I=\R$ there are no bounded functions in $C^\infty_0(\R)$ such that
$\eta$ can be defined as in eq.(\ref{etas1}).
However let ${\{a_n\}}_{n\in\N}\; ,a_{n+1}>a_n>0$ be a sequence in $\R$
with $a_n\ra\infty$ as $n\ra\infty$. Define $U_n=\; ]-a_n, a_n[$ and choose
$f_n,g_n\in C^\infty_0(\R)$ such that $f_n(t)=1 ,\; t\in U_n$,
$\; |f_n(t)|\leq 1 ,\; t\nin U_n $ and $\partial_tg_n(t)=1 , \; t\in U_n $,
$\; |\partial_tg_n(t)|\leq 1 ,\; t\nin U_n$. For $\eta_n=f_n\delta g_n$ it is
\be
\begin{array}{rcl}
(\Psi, (\pi_D(\eta_n)-\gamma^0)\Psi) & = & \int_{\R} dt (f_n\partial_tg_n-1)
{(\Psi, \gamma^0\Psi)}_s\\
 & & \\
 & \leq &
\int_{\R} dt{(\Psi,\Psi)}_s- \int_{-a_n}^{a_n} dt{(\Psi,\Psi)}_s
\end{array}
\;\;\;\; \Psi\in\cH\;\;.
\ee
Thus $\pi_D(\eta_n)$ converges in the strong operator topology to $\gamma^0$.
\qed

If $I=\R$ we add a formal limit point $\eta$ of the sequence $\eta_n$ to
$\Omega^1\cA$
with
\be
\pi_D(\eta)=\gamma^0\;\; .
\ee
This element is formal since we have not specified a topology
on $\OD$ which would allow to consider convergence of $\eta_n$ in $\OD$.
However,  except for the definition of the map $T$,
the element $\eta$ will appear only as an argument of $\pi_D$
or $\pds$ and therefore the limit is well defined in $\cB$. Furthermore we
note that
\be
\pds(\eta)= 0\;\; .
\ee

\begin{lemm}\label{le1}
For $\omega \in \Omega^k\cA$ it is
\be
\pi_D(\omega)=\pds(\omega) + \pi_D(\alpha)\;\; ,\;\;
\alpha\in\ker \pds \cap\Omega^k\cA\;\; .
\label{lemm1}
\ee
\end{lemm}

{\bf Proof:}
We prove this lemma by defining an algebra homomorphism $T$ on $\OA$ which is
an projection, i.e.~$T^2=T$. $\OA$ is generated by the zeroth and first
degree and therefore it is sufficient to define $T$ on those spaces.
Since $\cA=C^\infty(I,\cAs)$ it is $\partial_t A\in\cA\; \forall A\in\cA$.
We use this and the element $\eta$ to define
\be
\begin{array}{rclc}
T(A)       & =& A&\;\; A\in\cA\\
           &  &  &            \\
T(\delta A)& =&\delta A-\partial_tA\eta\; ;\;\; T(\eta) = 0 &\;\; A\in\cA\; .
\end{array}
\ee
For an arbitrary degree $k>1$ one obtains
\be
\begin{array}{rcl}
T & : & \Omega^k\cA \lra \Omega^k\cA\\
   &   & \\
T(A_0\delta A_1\cdots\delta A_k) & = & A_0 T(\delta A_1)\cdots T(\delta A_k)
\end{array}
\ee

Since
\be
\pi_D(\delta A) = [D_s,\pi(A)] + [D_t,\pi(A)]
= \pds(\delta A) + \pi_D(\partial_t A \eta) \label{tk1}
\ee
this map has the useful property that for any
$\omega \in \OA$
\be
\begin{array}{rcl}
\pds(\omega)  & = & \pi_D(T(\omega))\\
              &   &                 \\
\pds((1-T)\omega) & = & 0
\end{array}
\ee
It is $\pi_D(\Omega^0\cA)=\pds(\Omega^0\cA)=\pi(\cA)$ and for any $k$ it is
\be
\pi_D(\omega)=\pi_D(T(\omega)) + \pi_D((1-T)\omega)=\pds(\omega)+\pi_D(\alpha)
\; ,\;\;\forall \omega \in \Omega^k\cA
\ee
with $\alpha=(1-T)(\omega)\in\ker\pds$ and the lemma is proved.
\qed

\begin{lemm}\label{le2}
It is
\be
\pi_D(\cA) \subset \pi_D(\cJ^2_D) \subset \pi_D(\Omega^2\cA)
\ee
and thus there is a filtration on $\pi_D(\OA)$:
\be
\begin{array}{ccccccc}
\pi_D(\Omega^0\cA) & \subset & \pi_D(\Omega^2\cA) & \subset &
\pi_D(\Omega^4\cA) & \subset & \cdots\\
 & & & & & & \\
\pi_D(\Omega^1\cA) & \subset & \pi_D(\Omega^3\cA) & \subset &
\pi_D(\Omega^5\cA) & \subset & \cdots \;\;\; .
\end{array}
\ee
\end{lemm}

{\bf Proof:} Let us consider $\alpha =(f\otimes 1)\delta(g\otimes 1)
+(g\otimes 1)\delta(f\otimes1)-\delta(fg\otimes 1) \in \Omega^1\cA$. It is
\be
\pi_D(\alpha) = 0
\ee
but
\be
\pi_D(\delta\alpha)=2[D_t,\pi_t(f)][D_t,\pi_t(g)]=
2{\gamma^0}^2\partial_t f\partial_tg
=\pi(N)\pi(\partial_tf\partial_tg) \in \pi(\cA)\;\; ,
\ee
i.e. $\delta\alpha \in \cJ^2_D$ and with such elements all of $\pi(\cA)$
can be generated. Thus the lemma is proved. \qed

Suppose $\omega\in \ker\pi_D\cap\Omega^k\cA$. Lemma \ref{le1} allows
us to write
\be
0=\pi_D(\omega) = \pds(\omega) + \pi_D(\alpha)\;\; .
\ee
Because of lemma \ref{le2} we cannot infer that  $\pi_D(\omega)=0$ implies
$\pds(\omega) =0$. Thus if we would divide $\OA$ by the differential
ideal associated to $\pds$ in an analogous way as for $\pi_D$ in
eq.(\ref{jdef}) it may happen that the resulting differential algabra
contains elements which are not elements of $\OD\cA$ and therefore it
is not a subalgebra of $\OD\cA$. The correct differential ideal, which
leads to a graded subalgebra of $\OD\cA$ is constructed with the help
of the following ideal in $\OA$
\be
\begin{array}{rcl}
\cK^{2k} & = & \{\omega\in\Omega^{2k}\cA\; | \;\exists \alpha \in
\bigoplus^{k-1}_{j=0} \Omega^{2j}\cA\; ,\; \pds(\omega+\alpha)=0\} \\
  &  &  \\
\cK^{2k+1} & = & \{\omega\in\Omega^{2k+1}\cA\; | \;\exists \alpha \in
\bigoplus^{k-1}_{j=0} \Omega^{2j+1}\cA\; ,\; \pds(\omega+\alpha)=0\}\\
  &  &  \\
\cK & = & \bigoplus_{k=1}^\infty \cK^k\;\; .
\end{array}
\ee
Let us also define the ideal $\cK_0$
\be
\begin{array}{rcl}
\cK^k_0 & = & \ker\pds\cap\Omega^k\cA \\
  &  &  \\
\cK_0 & = & \bigoplus_{k=1}^\infty \cK_0^k\;\; .
\end{array}
\ee
A two sided differential ideal $\cN$ is obtained as in eq.(\ref{jdef})
by including the image of $\delta$ on $\cK$:
\be
\begin{array}{rcl}
\cN^k &=& \cK^k + \delta \cK^{k-1}\\
      & &                     \\
\cN & = & \bigoplus_{k=1}^\infty \cN^k \;\; .
\end{array}
\ee
The corresponding graded differential algebra $\ON\cA$ is then defined as
\be
\begin{array}{rcl}
{\ON}^k\cA &=& {\Omega^k\cA \over \cN^k}\\
      & &                     \\
\ON & = & \bigoplus_{k=0}^\infty {\ON}^k\;\; .
\end{array}
\ee
Let us denote by $\sigma_{\cN}$ the map on the quotient space
\be
\sigma_{\cN}:\pds(\Omega^k\cA)\lra\OD^k\cA\;\; .
\ee

The relation of $\OD\cA$ and $\ON\cA$ is determined by the relation of
$\cN$ and $\cJ$ and therefore it is useful to prove the following lemma
\begin{lemm}\label{le3}
It is
\be
\cK^k =(\ker\pi_D\cap\Omega^k\cA) \cup \cK_0^k\label{le3a}
\ee
and
\be
\cN^k=\cJ^k \cup \cK_0^k\;\; .\label{le3b}
\ee
\end{lemm}
{\bf Proof:}
It is clear that $(\ker\pi_D\cap\Omega^k\cA)\cup \cK_0^k \subset \cK^k$. Thus
we have to consider elements $\omega\in\cK^{2k}$ with
\be
0\neq \pds(\omega) =\sum_{j=0}^{k-1}\pds(\omega_{2j})\;\; ,\;\;
\omega_{2j}\in\Omega^{2j}\cA\;\; .
\ee
Because of lemma \ref{le1} there are $\alpha_{2j}\in\cK_0^{2j}$ and
$\alpha\in\cK_0^{2k}$ with
\be
\begin{array}{rcl}
\pi_D(\omega_{2j}-\alpha_{2j}) & = & \pds(\omega_{2j})\;\; ,\\
                               &   &                       \\
\pi_D(\omega-\alpha) & = & \pds(\omega)\;\; .
\end{array}
\ee
We define $\omega^\prime\in\Omega^{2k}\cA$ as
\be
\omega^\prime= \omega-\alpha -
\sum_{j=0}^{k-1} {(N^{-1}\eta)}^{2(k-j)}(\omega_{2j}-\alpha_{2j})\;\; .
\ee
Since
\be
\pds(\omega-\omega^\prime)=0
\ee
and
\be
\pi_D(\omega^\prime)=0
\ee
we infer that $\omega\in (\ker\pi_D\cap\Omega^{2k}\cA)\cup\cK_0^{2k}$. The
same is true for $\omega\in\cK^{2k+1}$ and therefore eq.(\ref{le3a}) is proved.

For the second part of the proof we compute $[\delta ,T]$:
\be
\begin{array}{rcl}
\delta T (A_0\delta A_1\cdots \delta A_k) & = &
\delta(A_0 T(\delta A_1)\cdots T(\delta A_k))\\
 & & \\
 & = &\delta A_0T(\delta A_1)\cdots T(\delta A_k) \\
 & & \\
 & & + \sum_{j=1}^kA_0T(\delta A_1)\cdots
(\delta A_j +{(-1)}^j \delta(\partial_t A_j \eta))\cdots
T(\delta A_k)\;\; ,\\
 & & \\
T\delta (A_0\delta A_1\cdots\delta A_k) &= &T(\delta A_0)
T(\delta A_1)\cdots T(\delta A_k)\;\;.
\end{array}
\ee
Thus
\be
\begin{array}{rcl}
[\delta ,T](A_0\delta A_1\cdots \delta A_k) & = &
\partial_tA_0\eta T(\delta A_1)\cdots T(\delta A_k)\\
 & & \\
 & & + \sum_{j=1}^kA_0T(\delta A_1)\cdots
(\delta A_j +{(-1)}^j \delta(\partial_t A_j \eta))\cdots T(\delta A_k) .
\end{array}
\ee
Therefore, with $\pds(\delta\eta)=0$, we conclude that for any
$\omega\in\cK_0$
\be
\pds([\delta ,T](\omega))=0\;\; .\label{dtcommu}
\ee
Furthermore it is for $\omega\in \cK_0$
\be
0=\pds(\omega)=\pi_D(T\omega)
\ee
and therefore
\be
\pi_D(\delta T\omega) \in \pi_D(\cJ)\;\; .
\ee
On the other hand it is
\be
\pds(\delta\omega)=\pi_D(T\delta\Omega)\;\; .
\ee
Together with eq.(\ref{dtcommu}) this proves eq.(\ref{le3b}). \qed

We now state the main result of this section which shows that $\ON\cA$ is
the ``space-like'' part of $\OD\cA$ in the sense that there is a ``time''
differential $d_t$ and a ``time-like'' differential one-form $dt$ in
$\OD\cA$. We then denote by ``space-like'' forms such elements in $\OD\cA$
which do not contain $dt$.
\begin{theo}
There is an element $dt\in\OD^1\cA$ such that for any $k$
\be
dt\omega-{(-1)}^k\omega dt=0\;\; \forall\omega\in\OD^k \label{dtc}
\ee
and
\be
\OD^k\cA = \ON^k\cA\; \oplus \; \ON^{k-1}\cA dt\;\; .
\ee
The differential $d$ on $\OD\cA$ is given as a sum of the two differentials
$d_s$ and $d_t$:
\be
d  =  d_s + d_t\;\; ,
\ee
\be
d_t(\sigma_D\pi_D(\omega))=
{(-1)}^k\sigma_D\pi_D(T(\partial_t\omega))dt
\;\; \omega\in\Omega^k\cA \label{dtdeff}
\ee
with
\be
\begin{array}{rcl}
\partial_t (A_0\delta A_1\cdots \delta A_k) & = &\partial_t A_0\delta A_1
\cdots \delta A_k\\
 & &        \\
 & & + \sum_{j=1}^k {(-1)}^{k-j}A_0\delta A_1\cdots \delta(\partial_tA_j)
\cdots\delta A_k\; .
\end{array}
\ee
\end{theo}

{\bf Proof:}\newline
{}From lemma \ref{le1} we know that $\pds(\OA)$ is a subalgebra of
$\pi_D(\OA)=\pds(\OA)\cup\pi_D(\cK_0)$
and hence
\be
\bigoplus_{k=0}^\infty {\pds(\Omega^k\cA)\over \pi_D(\cJ^k)}\subset\OD\cA
\ee
is a subalgebra of $\OD\cA$. Because of eq.(\ref{le3b}) we can conclude
that
\be
\bigoplus_{k=0}^\infty {\pds(\Omega^k\cA)\over \pi_D(\cJ^k)}=
\bigoplus_{k=0}^\infty {\pds(\Omega^k\cA)\over \pi_D(\cN^k)}= \ON\cA\;\; .
\ee
{}From eq.(\ref{le3a}) we infer that
\be
{\pds(\Omega^k\cA)\over \pi_D(\cJ^k)}\cap {\pi_D(\cK)\over\pi_D(\cJ^k)}
=\{ 0 \}\;\; .
\ee
Thus we can decompose $\OD\cA$ as follows
\be
\OD^k\cA=\ON^k\cA \oplus {\pi_D(\cK_0^k)\over \pi_D(\cJ^k)}\;\; .
\ee
We proceed by identifying $dt$ as
\be
dt=\sigma_D\pi_D(\eta)\;\; .\label{dtdef2}
\ee
Because of lemma \ref{le2} we know that $\eta^2\in\cJ^2$ and hence $dt^2=0$.
For any \newline
$A_0\delta A_1\cdots A_k\in\cK_0^k$ it is
\be
\begin{array}{rcl}
\pi_D(A_0\delta A_1\cdots\delta A_k) & = &
\pi_D(A_0)([D_s,\pi_D(A_1)] +\pi_D(\partial_tA_1)\pi_D(\eta))\cdots\\
 & & \\
 & &\cdots ([D_s,\pi_D(A_k)] +\pi_D(\partial_tA_k)\pi_D(\eta)) \\
 & &\\
 &=&\sum_{j=1}^k {(-1)}^{k-j}\pds(A_0 \delta A_1\cdots\partial_tA_j
 \cdots \delta A_k)\pi_D(\eta)+\pi_D(\alpha)
\end{array}\label{dt+}
\ee
where $\pi_D(\alpha)\in\cJ^k$ denotes the sum of terms with a factor
$\pi_D(\eta)^k ,\; k>1$. We also used property {\bf v.} of $D$ to anticommute
$\pi_D(\eta)$ to the left. From eq.(\ref{dt+}) we infer that
\be
{\pi_D(\cK^k)\over \pi_D(\cJ^k)} = \ON^{k-1}\cA dt
\ee
Eq.(\ref{dtc}) is also a consequence of property {\bf v.} of $D$.

Since $\OD\cA$ is generated by $\OD^1\cA$ and $dt^2=0$ it is sufficient
to show eq.(\ref{dtdeff})  for all $\mu\in \OD^1\cA$. For any $w\in \ON^1\cA$
let $A_0\delta A_1\in \Omega^1\cA$ be a representative, i.e.
\be
\sigma_{\cN}\pds(A_0\delta A_1)= w
\ee
Let us first compute the action of $d_s$ on $\sigma_D\pi_D(T(A_0\delta A_1))$,
which is the image of $w$ in $\OD\cA$
\be
d_s\sigma_D\pi_D(T(A_0\delta A_1))=\sigma_D\pi_D(T\delta A_0\delta A_1))=
\sigma_D([D_s,\pi(A_0)][D_s,\pi(A_1)])\;\; .
\ee
We use this to compute the action of $d$ on $\sigma_D(A_0[D_s,\pi(A_1)])$
\be
\begin{array}{rcl}
d\sigma_D\pi_D(A_0\delta A_1)&=&
\sigma_D\pi_D(\delta (A_0T(\delta A_1)))\\
 & & \\
 & = & \sigma_D(([D_s,\pi(A_0)]+[D_t,\pi(A_0)])\pi_D(T(\delta A_1)))
+\sigma_D\pi_D(A_0\delta T(\partial_t \eta))\\
 & & \\
 &=&d_s\sigma_D\pi_D(A_0\delta A_1)\\
 & & + dt\sigma\pi_D(\partial_tA_0T(\delta A_1))+
\sigma_D\pi_D(T(A_0\delta(\partial_tA_1)))dt\;\; .
\end{array}
\ee
This shows that
\be
d_t(\sigma_D\pi_D(A_0\delta A_1))=
(d-d_s)\sigma_D\pi_D(T(A_0\delta A_1))
=\sigma_D\pi_D(\partial_t(A_0\delta A_1))
\ee
and the theorem is proved.
\qed
\section{The Inner Product on $\OD\cA$ and the Lorentz Metric}\label{mink}
So far we have constructed a generalized differential algebra where we were
able
to identify the ``space-like'' and the ``time-like'' part because of the
structure $\cA = C^\infty(I,\cAs)$ of the algebra and the special form of the
k-cycle $(\cH, D)$. Following the lines presented by A.~Connes and J.~Lott
in \cite{colo,cobuch} it is now straightforward to construct a covariant
connection and curvature. However, there is still one important ingredient
missing which is neccessary to define an action or a Lagrange function
resp.~a Hamilton function, the objects we are interested in. In conventional
geometry one obtains an action or Lagrange function by integration over
appropriate differential forms. In \cite{co} A.~Connes showed that the correct
substitute for integration in non-commutative geometry is the Dixmier trace.
It is this trace which is used in the definition of actions in
\cite{colo,chams1,chams2}. However, we want to derive a
Hamilton function and therefore we do not have to integrate over the
non-commutative ``space-time'' but we have to integrate over a ``space-like''
surface. As before we will use the additional structure of $\cA$ and $(\cH, D)$
to define the correct operator theoretic substitute for integration
on ``space-like'' surfaces which will be the Dixmier trace on $\cHs$.

Let us first briefly recall the defintion of the Dixmier trace and some general
results about the inner product on $\OD\cA$ defined via Dixmier trace.
For a detailed account we refer to \cite{cobuch,Kbuch,GBV}.

The Dixmier trace \cite{dix} is the unique extension of the ususal trace to
the class $\cL^{(1,\infty)}(\cH)$ which is an ideal in the algebra of bounded
operators. The elements of this ideal are characterized by the condition
that for any $T\in\cL^{(1,\infty)}(\cH)$ the ordered eigenvalues $\lambda_i$
of $|T|$ satisfy
\be
\sup_{N}{1\over\log N}\sum_{i=0}^N \lambda_i < \infty\;\; .
\ee
On this ideal the Dixmier trace $Tr_\omega(\cdot )$ is defined as functional
with the property
\be
Tr_\omega(T) = \lim_{N\ra\infty}{1\over\log N}\sum_{i=0}^{N-1} \lambda_i\;\; .
\ee
If $\cA$ is an arbitrary subalgebra of a $C^*$-algebra with a finitely
summable k-cycle $(\cH, D)$ then $|D|^{-d}$ is in $\cL^{(1,\infty)}(\cH)$ for
some $d\in\N$, where $d$ corresponds to the dimension of the underlying
(non-commutative) space. Since
\be
Tr_\omega(|D|^{-d}) > 0
\ee
a inner product on $\pi_D(\OA)$ is obtained by defining for each $k$
\be
\begin{array}{rcl}
(\cdot,\cdot)^k&:&\pi_D(\Omega^k\cA)\times\pi_D(\Omega^k\cA)\lra\C\\
 & & \\
(\pi_D(\omega_1),\pi_D(\omega_2))^k&=&
Tr_\omega(\pi_D(\omega_1^*)\pi_D(\omega_2)|D|^{-d})\;\;
\omega_1,\omega_2\in \Omega^k\cA\;\; ,
\end{array}\label{topro}
\ee
which is positive if $\pi_D(\omega^*)=\pi_D(\omega)^*,\; \forall \omega\in\OA$.

Let us denote by $\cHp^k$ the Hilbert space completion of $\pi_D(\Omega^k\cA)$
and let $P^{(k)}$ be the orthogonal projection of $\cHp^k$ onto the orthogonal
complement of $\overline{\pi_D(\cJ^k)}\subset \cHp^k$ then an inner
product on $\OD\cA$ can be defined for each $k$ by
\be
\begin{array}{rcl}
<\cdot,\cdot>^k&:&\OD^k\cA\times\OD^k\cA\lra\C\\
 & & \\
<\sigma_D(W_1),\sigma_D(W_2)>^k&=& (P^{(k)}W_1,P^{(k)}W_2)\;\; ,
W_1,W_2\in\pi_D(\Omega^k\cA)\;\;,
\end{array}\label{toprof}
\ee
which is positive if $(\cdot ,\cdot )$ is positive.

This allows to identify $\OD^k\cA$ with a dense subspace of $\cHp^k$ and
hence there is a map
\be
\bc  : {\overline {\OD^k\cA}} \lra \cHp^k
\ee
with $Im(\bc) = {\overline {\pi(\cJ^k)}}^\perp$.

In the case, where $\cA=C^\infty(\cM)$ is the algebra of smooth functions on
a compact spin-manifold $\cM$ and $D=\dslash$ is the Dirac operator, $\OD\cA$
is the usual de Rham algebra \cite{cobuch} and the inner product is
\be
<w_1,w_2>=\int_{\cM} *w_1\wedge w_2   \;\; ,\;\; w_1,w_2\in\OD^k\cA
\ee
where $*w_1$ is the Hodge dual of $w_1$.

Let us now turn to our case where the algebra is of the form
$\cA=C^\infty(I,\cAs)$ where we would like to introduce a substitution for
integration on a space-like surface. However the ``space-like'' part of $\cA$
and $\OD\cA$ is characterized by $\cA_s$ and the smooth 1 parameter family
of k-cycles $(\cHs, D_s)$ over $\cAs$, which are finitely summable by
assumption. Therefore there is some $d$ (the dimension of the ``space-like''
part of the non-commutative space) such that for any $t\in I$ $\; |D_s|^{-d}$
is an operator on $\cH_s$ with $|D_s|^{-d}\in \cL^{(1,\infty)}(\cHs)$ and
\be
Tr_\omega(|D_s|^{-d})_s >0\;\; .
\ee
Here $Tr_\omega(\cdot)_s$ denotes the Dixmier trace on
$\cL^{(1,\infty)}(\cHs)$. Since for any $t\in I$ any \linebreak
$W\in\pi_D(\OD)$ is a bounded operator on $\cHs$ varying smoothly with $t$
\be
\begin{array}{rcl}
(\cdot,\cdot)^k_s&:&\pi_D(\Omega^k\cA)\times\pi_D(\Omega^k\cA)\lra\C^\infty(I)\\
 & & \\
(\pi_D(\omega_1),\pi_D(\omega_2))^k_s&=&
Tr_\omega(\pi_D(\omega_1^*)\pi_D(\omega_2)|D_s|^{-d})_s\;\; , \;\;
\omega_1,\omega_2\in\Omega^k\cA
\end{array}\label{spro}
\ee
defines a positive inner product on $\pi_D(\OA)$ for any $k$ and any (fixed)
$t\in I$ if \linebreak
$\pi_D(\omega^*)=\pi_D(\omega)^*,\;\forall \omega\in \OA$, a condition which
is met in our case (see {\bf vi.}), i.e.
\be
(W,W)_s =f(t)\geq 0\;\; ,\;\; \forall W\in\pi_D(\OA)\; ,\forall t\in I
\ee
With this inner product on $\pi_D(\OA)$ one can define an inner product on
$\OD\cA$ as in the general construction.
Let us denote by $\cHps^k$ the completion\footnote{We call a sequence
convergent with respect to $(\cdot ,\cdot)_s$ if it converges pointwise
for all $t\in I$.} of $\pi_D(\Omega^k\cA)$
with respect to $(\cdot,\cdot)_s$
and let $P^{(k)}_s$ be the orthogonal projection of $\cHps^k$ onto the
orthogonal complement of $\pi_D(\cJ^k)$ then for each $k$
\be
\begin{array}{rcl}
<\cdot,\cdot>^k_s&:&\OD^k\cA\times\OD^k\cA\lra\C^\infty(I)\\
 & & \\
<\sigma_D(W_1),\sigma_D(W_2)>^k_s&=& (P_s^{(k)}W_1,P_s^{(k)}W_2)_s\;\; ,
W_1,W_2\in\pi_D(\Omega^k\cA)
\end{array}\label{sprof}
\ee
defines a positive inner product on $\OD\cA$ for any $t\in I$. With this
product we will define Lagrange functions and the Hamilton formalism.
As in the general case there is a map
\be
\bc_s  : {\overline {\OD^k\cA}} \lra \cHps^k
\ee
and hence we can identify $\OD^k\cA$ with a dense subspace of $\cHps^k$.

The definition of the inner product in eq.(\ref{sprof}) allows for an
important freedom in the choice of the k-cycle over $\cA$, which deserves
some discussion. For any \linebreak
$w_1\in\ON^k\cA\subset\OD^k\cA$ and any
$w_2\in \ON^{k-1}\cA dt\subset\OD^k\cA$ it is
\be
<w_1,w_2>_s=\left(\bc_s(w_1),\bc_s(w_2)\right)_s=
Tr_\omega(\bc_s(w_1)^*\bc(w_2))_s= 0
\ee
since $\bc_s(w_1)^*\bc(w_2)$ contains an odd number of commutators with $D_s$
which are off diagonal (with respect to the block diagonal structure of
eq.(\ref{blockdia})). This proves the following lemma:
\begin{lemm}\label{ordeco}
The decomposition
\be
\OD^k\cA=\ON^k\cA \oplus \ON^{k-1}\cA dt
\ee
is orthogonal with respect to the inner product $<\cdot,\cdot>_s$.
\end{lemm}
We have seen that condition {\bf iii.} of $D$, i.e.~$D$ has a compact inverse
and \linebreak
$|D|^{-d}\in \cL^{(1,\infty)}(\cH)$, is crucial for the definition
of the inner products (\ref{topro}) and (\ref{toprof}). However, we will
restrict ourselves to ``integration on space-like'' surfaces and hence use
the inner products defined by eq.(\ref{spro}) and (\ref{sprof}). Here we
only need that $D_s$ has a compact inverse and that
$|D_d|^{-d}\in\cL^{(1,\infty)}(\cHs)$ for some $d$ which is guaranteed by
condition {\bf vi.}. We can use this freedom and change the definition of
$D_t$ by choosing $\gamma^0$ self-adjoint. As a consequence we find for
any element $\omega\in\OD\cA$
\be
\begin{array}{rcl}
{(d_s\omega)}^* &=& -d_s (\omega^*)\\
           & &           \\
{(d_t\omega)}^* &=& d_t (\omega^*)\; .
\end{array}\label{minkmet}
\ee
Following A.~Chamseddine et.al.~\cite{CFF} we introduce a
generalized metric on $\OD^1\cA$\footnote{Strictly speaking the metric is
introduced on $\overline{\OD^1\cA}$ which is the Hilbert-space completion of
$\OD^1\cA$. However, we assume that the construction holds on $\OD^1\cA$.}.
In this context the $\cA$-module $\OD^1\cA$
is interpreted as the generalized cotangent bundle over a \linebreak
non-commutative space. We define the metric
\be
g(\cdot,\cdot):\OD^1\cA\times\OD^1\cA\lra\cA
\ee
by the following equation
\be
<A,g(v,w)>_s=-{Tr_\omega(A^*\bc_s(v^*)\bc_s(w)\|D_s|^{-d})}_s
\;\; ,\;\; \forall A\in\cA ;\; v,w\in\OD^1\cA\;\; .\label{metdef}
\ee
This metric enjoys the property
\be
g(Av,Bw)=A^*g(v,w)B\;\; ,\;\; \forall A,B\in\cA ;\; v,w\in\OD^1\cA\;\; .
\ee
An important property of this metric is stated in the following theorem
\begin{theo}
If $\gamma^0$, as defined in eq.(\ref{etacon}), is self-adjoint, i.e.
\be
\gamma_0 = \gamma_0^\dagger
\ee
then $g(\cdot,\cdot)$, as defined in eq.(\ref{metdef}), is generalized
Minkowskian metric, which is positive definite on $\ON^1\cA$ and negative
definite on the $\cA$-module generated by $dt$.
\end{theo}

{\bf Proof:}
Applying the arguments presented in \cite{CFF} to our case, we conclude that
$g(\cdot,\cdot)$ defines a positive definite Riemannian metric on
$\ON^1\cA\subset\OD^1\cA$. From lemma~\ref{ordeco} we infer that
\be
g(v,dt)=0\;\; , \;\; \forall v\in\ON^1\cA\;\; .
\ee
{}From the definitions eq.(\ref{spro}), eq.(\ref{sprof}) and the definition
of $\gamma^0$ in eq.(\ref{etacon}) it follows that
\be
g(dt,dt)=-\gamma^0\gamma^0=-N\in\cA\label{timsign}
\ee
and the theorem is proved. \qed

This theorem completely justifies the
terminology of ``space-like'' and ``time-like'' since with the choice
${\gamma^0}^\dagger=\gamma^0$ it is possible to identify time like elements
of $\OD^1\cA$ as elements with negative norm, i.e.~elements $v\in\OD^1\cA$
with
\be
g(v,v)=|v|^2<0\;\; .
\ee
For the rest of this article we will keep the choice
${\gamma^0}^\dagger=\gamma^0$, which means we are working on a non-commutative
Minkowski space.

We end this section with some furhter definitions and some assumption on
the algebra $\cA$ which will be useful in the Hamiltonian framework.
The first definition is a slight generalization of eq.(\ref{metdef}).
We associate to any $v^{(l)} \in \OD^l\cA,\; l\geq 0$ a map $i_l(v^{(l)})$,
which is defined for all $k\geq 0$ by
\be
\begin{array}{rl}
<w_1,i_l(v^{(l)})w_2>_s=
Tr_\omega(\bc_s(w_1^*)\bc_s((v^{(l)})^*)\bc_s(w_2)|D|^{-d})_s
&=<vw_1,w_2>_s\\
 & \forall w_1\in\OD^k\cA ,\;\forall w_2\in\OD^{k+l}\cA\; .
\end{array}
\ee
This map is well defined as can be seen by applying the arguments presented
in \cite{CFF} for the definition of the metric. Thus we have defined a map
which decreases the degree of forms
\be
i_l(v^{(l)}) : \OD^{k+l}\cA \lra {\overline{\OD^{k}\cA}}\;\; .
\ee
For the second definition we have to make a furhter assumption on the algebra
$\cA$ and the k-cycle $(\cH, D)$ over $\cA$. Namely that for any
$v\in\OD^k\cA$, $k>0$, there is a $C_v\in\R$ such that for all
$w\in\OD^{k-1}\cA$
\be
|<v, dw>_s|^2\leq C_v <w,w>_s\label{regcon}
\ee
This condition is fulfilled, for example, if $\cA=C^\infty(M)$ and $D$ is
the Dirac-operator on $M$ or if $\cA$ is a finite dimensional algebra or
if $\cA$ is a product of the first two cases. Thus eq.(\ref{regcon}) is
fulfilled for the class of algebras which has been used for model building
in physics so far. This condition ensures that there is a well defined
adjoint operator $d_s^*$ of $d_s$ on $\OD^k\cA$
\be
d_s^* : \OD^{k}\cA \lra {\overline{\OD^{k-1}\cA}}
\ee
which is uniquely defined by
\be
<d_s^*v,w>_s:= <v,d_sw>_s\;\; \forall v\in \OD^k\cA,\; \forall w\in\OD^{k-1}\cA
\;\; .
\ee
Furhtermore we assume that the smooth 1-parameter family of k-cycles
$(\cHs, D_s)$ is tame \cite{GBV}, i.e.
\be
Tr_\omega( [W_1,W_2] |D_s|^{-d})= 0\;\; , \;\; W_1,W_2\in\pi_D(\OA)
\ee
and
\be
\begin{array}{rcl}
i_l(v^{(l)})(\OD\cA)&\subset&\OD\cA\\
            &       &      \\
d_s^*(\OD\cA) &\subset&\OD\cA
\end{array}\;\; .\label{gbm}
\ee
These conditions are fulfilled in the above mentioned examples.
\section{Lagrange and Hamilton Function for Yang-Mills Theory}
Now we have all basic objects at hand which are necessary to define
a Lagrange function and the corresponding Hamilton function for
Yang-Mills theory in non-commutative geometry. However, we start with
a brief exposition of Yang-Mills theory in non-commutative geometry
as it was introduced A.~Connes and J.~Lott \cite{colo,cobuch}, which
allows us to fix our notation. A comprehensive presentation of this
subject can be found in \cite{cobuch,Kbuch,GBV}.

Yang-Mills theory is formulated on vector bundles. In the algebraic language
a vector bundle is a finitely generated projective module over $\cA$ which
we denote by $\cE$. Any finitely generated module $\cE$ can be obtained from
a free module $\cE_0=\cA^N$ with the help of some idempotent
$e\in\cA^{N\times N}$,
which means that we we can write $\cE=e\cA^N$. In our case, the structure
of $\cA=C(I,\cAs)$ implies that $\cE = C(I, \cEs)$, where $\cEs$ is a finitely
generated projective module over $\cAs$.

Furthermore we need a Hermitian
structure on $\cE$,i.e., a sesquilinear form
\be
(\cdot,\cdot)_{\cE}:\cE\times\cE\lra \cA
\ee
with the following properties
\begin{itemize}
\item $(A\zeta,B\eta)_{\cE}=A^*(\zeta,\eta)_{\cE}B\;\; ,\;\;
\forall \zeta,\eta\in\cE ,\; A,B\in\cA$
\item $(\zeta,\zeta)_{\cE}\geq 0\;\; ,\;\; \forall\zeta\in\cE$
\item $\cE$ is self dual for $(\cdot,\cdot)_{\cE}$.
\end{itemize}
If we write $\cE=e\cA^n$ the hermitean structure requires that $e$ is
self-adjoint.

We extend $\cE$ to a right module $\cEt$ over $\OD\cA$
\be
\cEt^k=\cE\otimes_{\cA}\OD^k\cA\;\;\; ,\;\;\;
\cEt=\cE\otimes_{\cA}\OD\cA\;\;
\ee
and also
\be
(\cdot,\cdot)_{\cEt}:\cEt\times\cEt\lra \OD\cA\;\; .
\ee
A connection is defined as a linear map
\be
\nabla : \cEt^k\lra\cEt^{k+1}
\ee
such that
\be
\nabla(\zeta w)=\nabla(\zeta)w+(-1)^kdw\;\; ,\;\;
\zeta\in\cEt^k ,\; w\in\OD\cA\;\; .
\ee
One also requires that the connection is compatible with the metric
$(\cdot,\cdot)_{\cEt}$, which for Euklidean k-cycles, i.e., for
$D^\dagger=D$ is equivalent to the condition
\be
(\zeta,\nabla\eta)_{\cEt}-(\nabla\zeta,\eta)_{\cEt} = d(\zeta,\eta)_{\cEt}
\;\; ,\;\; \zeta,\eta\in\cE\;\; . \label{compcon}
\ee
The set of compatible connections form an affine space and for any two
compatible connections $\nabla ,\; \nabla^\prime$ it is
\be
\nabla-\nabla^\prime = \vA\in\Homa\;\; .
\ee
Note, that the definition of a compatible connection depends on the
definition of the $*$-operation on $\OA$ and the choice of $D$ for the
k-cycle over $\cA$. In our case we have $D_s^\dagger=D_s$ and
$D_t^\dagger = -D_t$. Thus condition (\ref{compcon}) is valid only on
the space-like part of the connection. For the time-like part of the connection
the compatibility condition reads
\be
(\zeta,\nabla_t\eta)_{\cEt}+(\nabla_t\zeta,\eta)_{\cEt} =
d_t(\zeta,\eta)_{\cEt}
\;\; ,\;\; \zeta,\eta\in\cE\;\; .
\ee

One can check (see e.g.\cite{GBV}) that for $\cE=e\cA^N$
\be
\nabla_0 \zeta=ed\zeta\;\; ,\;\; \zeta\in\cE
\ee
defines a compatible connection. Thus any compatible connection $\nabla$ can be
written as
\be
\nabla=\nabla_0+\vA\;\;, \;\; \vA\in\Homa\;\; .
\ee
Here we used that the restriction on $\nabla$ to $\cE$ already defines the
connection uniquely on $\cEt$. The curvature $\vF$ is obtained by taking the
square of the connection
\be
\vF=\nabla^2=e(de)^2 +ede \alpha e +ed\alpha e -e\alpha de+ e\alpha e\alpha
\in\Homb
\ee
with $e\alpha e= \vA$ and $\alpha \in \cA^{N\times N}\otimes_{\cA}\OD^1\cA$.

Connection and curvature transform covariantly under unitary transformations
\linebreak $U(\cE)=\{ u\in\mbox{End}_{\cA}(\cE)| uu^*=u^*u = 1\}$, i.e.
\be
\vF^\prime = u\vF u^*\;\;\; , \;\;\; \nabla^\prime = u\nabla u^*
\ee
from which we infer that the vector-potential $\vA$ transforms as follows
\be
\vA^\prime = u\vA u^* + udu^*\;\; .
\ee
The inner product on $\OD\cA$ and the hermitean structure on $\cE$ induce
a natural inner product on $\Homk$ for any $k$. We want to construct this
product explicitely and therefore we note that any $\diva{T}\in\Homk$ can be
written as
\be
\diva{T} = \sum_{r,s=1}^N e_{ik}w_{rs}e_{lj}\;\; ,\;\;  w_{kl}\in\OD^k\cA\;\;.
\ee
In this notation the inner product $(\cdot ,\cdot)$ on $\Homk$ can be defined
as
\be
(\cdot ,\cdot) : \Homk\times\Homk \lra C^\infty(I)
\ee
\be
\begin{array}{rcl}
(\diva{T}^{(1)},\diva{T}^{(2)}) &=&
tr Tr_\omega(\bc_s({\diva{T}^{(1)}}^\dagger)\bc_s(\diva{T}^{(2)})
|D_s|^{-d})_s\\
 & &\\
 &=&\sum_{j,k=1}^N \sum_{r,s=1}^N \sum_{p,q=1}^N
<e_{rj}w_{sr}^{(1)}e_{ks},e_{jp}w_{pq}^{(2)}e_{qk}>_s
\end{array}
\;\; ,\;\; w_{rs}^{(1)},w_{pq}^{(2)}\in\OD^k\cA,\label{hompro}
\ee
We use this inner product to define the
Lagrange function $L$ for Yang-Mills theory in non-commutative geometry:
\be
L(\vA)=-\frac{1}{4}(\vF ,\vF)\in C^\infty(I)
\ee
The action $S$ for Yang-Mills theory is obtained by integrating the Lagrange
function $L$ over time
\be
S(\vA)=\int_{t_1}^{t_2} dt L(\vA) = -\frac{1}{4}\int_{t_1}^{t_2} dt (\vF ,\vF)
\;\; .
\ee
So far we have discussed the general case where $\cE$ is a finitely generated
$\cA$-module. However, we now will restrict ourselves to the case where
$\cE=\cA^N$ is a free module. However, note that the formalism which will
be presented in the following can be generalized to finitely generated
$\cA$-modules. The reason for the restriction is just to avoid unecessary
complicated formulas.

Because of lemma \ref{ordeco} there is also an orthogonal decomposition of
$\Homk$ with respect to the inner product $(\cdot ,\cdot)$:
\be
\Homk=\Homks \oplus \Homkt
\ee
and therefore we can $\vF$ decompose as follows
\be
\begin{array}{rcl}
\vF &=& \vF_{st} + \vB\\
    & &             \\
\vF_{st} &=& d_t\vA_s + \nabla_s\vA_t\in\Homst\\
    & & \\
\vB &=& d_s\vA_s + \vA_s^2\in\Homss
\end{array}
\ee
where $\vA_s\in\Homs$ is the space-like part of $\vA$ and \newline
$\vA_t\in\Homt$ is
the time-like part of $\vA$. $\; \nabla_s= d_s +\vA_s$ denotes the space-like
part of the connection. With this decomposition $L$ becomes
\be
L=-\frac{1}{4}\left((\vF_{st},\vF_{st})+(\vB ,\vB)\right)\;\; ,
\ee
where the first term on the right hand side is positive and the second term
is negative.

Now we define the canonical momenta in the usual way, namely the variation of
$L$ with respect to the time derivative of the variables at some fixed time
$t$.
In our case we have to vary $L$ with respect to $d_t\vA$. We find that
\bes
\vE_s &=& {\delta L\over \delta dt\vA_s}=-{1\over 2}<\vF_{st},\cdot >
\in{\Homst}_t^\star
\label{esdef}\\
\vE_t &=& {\delta L\over \delta dt\vA_t}= 0\;\; .\label{etdef}
\ees
Here ${\Homst}_t^\star$ denotes the image of the map (at some fixed time $t$)
\be
\begin{array}{rcl}
\star   & : & {\cal T}_{st} \lra {\cal T}^\star_{st}\\
\star(T)& = & (T,\dot)\; ,\;\; T\in{\cal T}
\end{array}
\ee
restricted to $\Homst$, where ${\cal T}_{st}$ is the Hilbert space completion
of $\Homst$ and ${\cal T}^\star_{st}$ is the dual Hilbert space of
${\cal T}_{st}$. However, we use the map $\star^{-1}$ to identify
${\Homst}_t^\star$ with $\Homst_t$ and thus we consider the canonical
momentum $\cE$ as an element of $\Homst_t$. As in usual Yang-Mills theory
we see that there are no canonical momenta for $\vA_t$. Thus eq.(\ref{etdef})
are primary constraints.

We define the Hamiltonian $H$ as
\be
\begin{array}{rcl}
H &=& \vE(d_t\vA)-L=
\frac{1}{4}\left(-(\vE_s ,\vE_s) + (\vB ,\vB)\right) -
(\nabla_s^*\vE_s,\vA_t)\\
 & & \\
 & = &H_0 - \diva{G}(\vA_t)
\end{array}
\label{ncgham}
\ee
\be
H_0= \frac{1}{4}\left(-(\vE_s ,\vE_s) + (\vB ,\vB)\right)\;\; ,\;\;
\diva{G}=(\nabla_s^\star\vE_s,\vA_t)\;\; ,
\ee
where
\be
\nabla_s^*: {\Homk}\lra{\mbox{Hom}_{\cA}(\cE,\cE\otimes_{\cA}\OD^{k-1}\cA)}
\ee
is defined by
\be
(\diva{T}_1,\nabla_s\diva{T}_2)=(\nabla_s^*\diva{T}_1,\diva{T}_2)
 ,\;\;\diva{T}_1\in{\Homk},\;\diva{T}_2\in
\mbox{Hom}_{\cA}(\cE,\cE\otimes_{\cA}\OD^{k-1}\cA) .
\ee
Such a map exists because of the assumption (\ref{gbm}).
Note that $H_0$ is positive since it is $(\vE_s,\vE_s)\leq 0$.
As one may have expected, the Hamiltonian for Yang-Mills theory in
non-commutative geometry is formally exactly the same as for conventional
Yang-Mills theory. However, the Hamiltonian in eq.(\ref{ncgham})
is defined purely algebraic and therefore still makes sense in cases where
there is no space-time manifold.

\section{The Poisson Bracket and Time Evolution}
{}From the discussion of the previous section we infer that the canonical
phase-space $\Gamma_0$ of Yang-Mills theory in non-commutative geometry
is
\be
\Gamma_0\subset {\Homst}_t \oplus \Homs_t\;\; ,
\ee
where the subscript $t$ indicates that we have fixed the time $t$ when the
momenta were defined. Thus the elements of the phase-space $\Gamma_0$ do not
have any time dependence. More generally, we define for any $k$
\be
{\Homk}_t= {\Homk \over {\cal I}^k_t}
\ee
where ${\cal I}_t$ is the graded ideal
\be
{\cal I}_t=\{ \diva{z}\in\Hom | \diva{z}(t)= 0\}\;\; .
\ee
Since from now on all objects are considered at some
fixed time $t$ we drop the subscript $t$ in order to simplify notation.

However, there are some restrictions on the elements of $\Gamma_0$. The
first one is a reality constraint on the variables which originates from
the condition that $\vA$ is a compatible connection, i.e.
\be
\vA^\dagger = \vA\;\; .
\ee
Since
\be
\vE=-d_t\vA -\nabla_s\vA_t
\ee
the compatibility condition on $\vA$ implies that
\be
\vE^\dagger = -\vE
\ee
Thus the canonical phase-space of Yang-Mills theory in non-commutative
geometry is
\be
\Gamma_0 =\{(\vA,\vE)\in{\Homst}_t \oplus \Homs_t |
(\vA^\dagger,\vE^\dagger)=(\vA,-\vE)\} \; .\label{phase0}
\ee

In eq.(\ref{phase0}) we also have used the fact that there is no
canonical momentum for $\diva{A}_t$ and hence this variable plays the role of
a Lagrange multiplier. Thus we can read off from eq.(\ref{ncgham}) the
secondary constraint on the elements $\vA,\; \vE$ of $\Gamma_0$
(we suppressed the index $s$), namely
\be
\diva{G}(\vA_t)=(\nabla^*\vE,\vA_t)=0\; ,\;\;
\forall \vA_t \in\Homt\;\; . \label{Gauss}
\ee
This is the Gau\ss-Law in non-commutative geometry.

However, we have not defined a Poisson bracket for this space so far. A
Poisson bracket is a antisymmetric linear map $\{\cdot ,\cdot\}$ on a
suitable space of functions $\cC$ on $\Gamma_0$. Therefore we first
have to define $\cC$.

We take for $\cC$ the algebra of functions on $\Gamma_0$ which contain
arbitrary finite powers of the elements $\vA, \vE\in\Gamma_0$ and their
derivatives (of finite order). For any $w\in\OD\cA$ we define
\be
\begin{array}{rclcrcl}
w^{(2k,0)}w &:=& (d_s^*d_s)^k w &\;\; ,\;\; &
w^{(0,2k)}w &:=& (d_sd_s^*)^k w \;\; , \\
 & & & & & & \\
w^{(2k+1,0)}w &:=& (d_sd_s^*)^k d_s w &\;\; ,\;\; &
w^{(0,2k+1)}w &:=& (d_s^*d_s)^k d_s^* w \;\; .
\end{array}
\ee
General combinations of derivatives are denoted by
$w^{(k,l)}=w^{(k,0)}+w^{(0,l)},\; w\in\OD\cA$.
Those elements are well defined because of assumption (\ref{gbm}).

Furhtermore we need the analogue of partial integration in non-commutative
geometry. For this purpose we define for any $k\geq 0$ the map $pr_k$
\be
pr_k : \pi_D(\OA) \lra \OD^k\cA
\ee
by the equation
\be
<v,{pr}_k(W)>_s = Tr_\omega(\bc_s(v)^* W |D_s|^{-d})_s\;\; ,\;\;
\forall v\in\OD^k\cA\; , W\in\pi_D(\OA)\;\; .
\ee
Again assumption (\ref{gbm}) ensures that this map exists. With the help
of this map we can define the analogue of partial integration for all
$v\in\OD^k\cA, W\in\bc_s(\OD\cA)$ by
\be
\begin{array}{rcl}
Tr_\omega(\bc_s(d_sv) W |D_s|^{-d})_s &=& -Tr_\omega(\bc_s(v)
\bc_s(d^*_s{pr}_{k+1}(W))|D_s|^{-d})_s\\
 & & \\
Tr_\omega(\bc_s(d_s^*v) W |D_s|^{-d})_s &=& -Tr_\omega(\bc_s(v)
\bc_s(d_s{pr}_{k-1}(W))|D_s|^{-d})_s
\end{array}\; .
\ee

It is convenient to consider the subalgebra $\cP(\Gamma_0)$ of the
algebra of continuous maps from $\Gamma_0$ to
$\mbox{Hom}_{\cA}(\cE,\cE\otimes_{\cA}\pi_D(\OA))$, which is generated by
elements $P_m^{(j,k)}$ of the form
\be
P_m^{(j,k)}=\bc_s({pr}_m(\bc_s(\vz_1)\cdots\bc_s(\vz_n))^{(j,k)})\;\;, \;\;
j,k\geq 0, n>0
\ee
with
\be
\vz_l\in\{ \vA,\vE,\; N_0, N_s, N_t\}\;\; ,\;\; (\vA,\vE)\in\Gamma_0\; .
\ee
The elements $N, N_s, N_t$ with
\be
\begin{array}{rclc}
N &\in&{\mbox{Hom}_{\cA}(\cE,\cE)} ,\;& N^\dagger=-N\\
 & & & \\
N_s &\in&\Homs ,\;& N_s^\dagger = N_s\\
 & & & \\
N_t &\in&\Homt , \;& N_t^\dagger=-N_t
\end{array}
\ee
play the role of test functions.
We obtain $\cC\subset C(\Gamma_0, \C)$ by taking the trace of the elements
in $\cP$
\be
\cC:=\{ F\in C(\Gamma_0,\C)| F= tr Tr_\omega(P {|D_s|}^{-d})_s,\;P\in \cP\}\; .
\ee
Having specified the space of functions on $\Gamma_0$ we define the Poisson
bracket $\{\cdot,\cdot\}$  by the following set of rules
\be
\begin{array}{rcl}
\{ tr Tr_\omega(P_1|D_s|^{-d})_s , tr Tr_\omega(P_2 |D_s|^{-d})_s\}& =&
tr Tr_\omega( G(P_1,P_2) |D_s|^{-d})_s\\
 & & \\
 &=& - tr Tr\omega( G(P_2,P_1) |D_s|^{-d})_s\;\; ,
\end{array}
\ee
The functional $G(\cdot ,\cdot)$ is the non-commutative generalization of the
$\delta$-distribution.

For any $P_p,P^\prime_q\in\cP,\; 1\leq p\leq k,\; 1\leq q\leq l$ it is
\be
\begin{array}{ll}
trTr_\omega(G(P_1\cdots P_k,P_1^\prime\cdots P_l^\prime)|D_s|^{-d})_s=& \\
 & \\
=\sum_{cp_k}\sum_{cp_l}
trTr_\omega(P_{cp_k(1)}\cdots P_{cp_k(k-1)}G(P_{cp_k(k)},P^\prime_{cp_l(1)})
P^\prime_{cp_l(2)}\cdots P^\prime_{cp_l(l)})|D_s|^{-d})_s\; , &
\end{array}
\ee
where $\sum_{cp_k}$ denotes the sum over the cyclic permutations of the first
$k$ indices and $\sum_{cp_l}$ denotes the sum over the cyclic permutations of
the last $l$ indices.

For any $\bc_s(d_sv)\in \cP,\; v\in\OD^k\cA$ and for any
$\bc_s(d^*_sv)\in \cP,\; v\in\OD^k\cA$ it is $\forall P_1,P_2\in\cP$
\be
\begin{array}{rcl}
trTr_\omega(G(P_1,\bc_s(d_sv)) P_2 |D_s|^{-d})_s &=&
-trTr_\omega(G(P_1,\bc_s(v))\bc_s(d^*_s{pr}_{k+1}(P_2))|D_s|^{-d})_s\\
 & & \\
trTr_\omega(G(P_1,\bc_s(d_s^*v)) P_2 |D_s|^{-d})_s &=&
-trTr_\omega(G(P_1,\bc_s(v)) \bc_s(d_s{pr}_{k-1}(P_2))|D_s|^{-d})_s
\end{array}\; .
\ee
And finally we define for the basic fields
$\vz_1 ,\vz_2 \in\{ \vA,\vE,\; N_0, N_s, N_t\}$
\be
\!tr Tr_\omega( P_1 G(\bc_s(\vz_1),\bc_s(\vz_2)) P_2 |D_s|^{-d})_s =
\left\{
\begin{array}{lcr}
\!\! tr Tr_\omega(P_1{\gamma^0}^{-1}{id}_{\cE}P_2|D_s|^{-d})_s &\mbox{if}&
\vz_1=\vA ,\vz_2=\vE\\
 & & \\
\!\! -tr Tr_\omega(P_1{\gamma^0}^{-1}{id}_{\cE}P_2|D_s|^{-d})_s &\mbox{if}&
\vz_1=\vE ,\vz_2=\vA\\
 & & \\
 0 & &\mbox{otherwise}
\end{array}\right.
\ee
This completes the definition of the phase-space and the Poisson algebra.

The time evolution of the system is determined by the Hamiltonian $H$. For
any element $F\in\cC$ it is
\be
\dot{F} = \{ F, H\}
\ee
where the dot denotes the time derivative of $F$. However, the Hamiltonian
is not uniquely defined for this system since for some arbitrary
$\Lambda \in\mbox{Hom}_{\cA}(\cE,\cE)$ we can add $\diva{G}(\Lambda dt)$
to the Hamiltonian without changing physics. This is possible because
$\diva{G}(\Lambda)dt$ has to vanish on the physical subspace of $\Gamma_0$.
Furhtermore, consistency requires that the condition eq.(\ref{Gauss}) is
time-independent which leads to the following equations
\bes
\{\diva{G}(\Lambda dt), H_0\} &\approx& 0\; ,\label{GH}\\
\{\diva{G}(\Lambda_1 dt),\diva{G}(\Lambda_2 dt)\} &\approx& 0\;\; .\label{GG}
\ees
Here $\approx$ means that the equtions hold modulo constraints
This implies that the constraints have to form a closed algebra.

Let us check that eqs.(\ref{GH},\ref{GG}) are satisfied. We start with
eq.(\ref{GG}):
\be
\begin{array}{rcl}
\{\diva{G}(\Lambda_1 dt),\diva{G}(\Lambda_2 dt)\} &=&
trTr_\omega(G(
(\bc_s(\nabla_s\Lambda_1)\gamma^0\bc_s(\vE),
(\bc_s(\nabla_s\Lambda_2)\gamma^0\bc_s(\vE))|D_s|^{-d})_s\\
 & & \\
&=&-trTr_\omega(
\bc_s(\nabla_s\Lambda_1)
(\Lambda_2\gamma^0\bc_s(\vE) - \gamma^0\bc_s(\vE)\Lambda_2)
|D_s|^{-d})_s\\
 & & \\
& &+trTr_\omega(
(\Lambda_1\gamma^0\bc_s(\vE) - \gamma^0\bc_s(\vE)\Lambda_1)
\bc_s(\nabla_s\Lambda_2)
|D_s|^{-d})_s\\
 & &\\
 &=&-\diva{G}((\Lambda_1\Lambda_2-\Lambda_2\Lambda_1)dt)
\end{array}
\ee
Before we turn to eq.(\ref{GH}) it is useful to compute the following
bracket
\be
\begin{array}{rcl}
trTr_\omega
(\Lambda \gamma^0 G(\bc_s(\vE), {1\over 2} \bc_s(\diva{B})^2)
|D_s|^{-d})_s
 &=& -trTr_\omega(\Lambda \bc_s(d_s^*\diva{B}))|D_s|^{-d})_s\\
 & & \\
 & & +trTr_\omega(\bc_s(\vA)\bc_s(\diva{B})
-\bc_s(\diva{B})\bc_s(\vA))|D_s|^{-d})_s\\
 & & \\
 &=& -trTr_\omega(\Lambda \bc_s(\nabla_s^*\diva{B})|D_s|^{-d})_s
\end{array}\label{rotB}
\ee
If we now insert $\Lambda=\bc_s(\nabla_s\Lambda_0)$ in eq.(\ref{rotB})
we obtain
\be
\{\diva{G}(\Lambda_0 dt), {1\over
2}trTr_\omega(\bc_s(\diva{B})^2|D_s|^{-d})_s\}
=-trTr_\omega((\bc_s(\nabla_s\Lambda)\bc_s(\nabla_s^*\diva{B})|D_s|^{-d})_s=0 .
\ee
The remaining part is
\be
\{\diva{G}(\Lambda_0 dt), {1\over 2}trTr_\omega(\bc_s(\vE)^2|D_s|^{-d})_s\} =
trTr_\omega(\Lambda_0\bc_s(\vE)^2-\bc_s(\vE)^2\Lambda_0|D_s|^{-d})_s = 0\; .
\ee
Hence the conditions eq.(\ref{GH}) and eq.(\ref{GG}) are fulfilled and
the constraints $\diva{G}(\Lambda dt)$ form a complete set of first-class
constraints generating the symmetry of the theory. Thus the observables of
the theory are elements $F\in \cC$ with
\be
\{\diva{G}(\Lambda dt), F\} = 0\;\; .
\ee

The time evolution of the basic fields $\vA, \vE$ can be computed by
considering
\be
\{ trTr_\omega(\bc_s(\Lambda)\bc_s(\vA)|D_s|^{-d})_s, H_0\}=
trTr_\omega(\bc_s(\Lambda){(\gamma^0)}^{-1}\bc_s(\vE)|D_s|^{-d})_s
\ee
{}From this and eq.(\ref{rotB}) we infer that the time evolution of the
basic fields is (modulo gauge transformations)
\be
\begin{array}{rcl}
\dot{\vA} &=& -pr_1({(\gamma^0)}^{-1}\bc_s(\vE))\\
 & & \\
\dot{\vE} &=& -pr_2{(\gamma^0})^{-1}\nabla_s^*\diva{B}\;\; .
\end{array}
\ee
Equivalently, with $\vE=\vE_0dt$, we can write
\be
\begin{array}{rcl}
\dot{\vA} &=& \vE_0\\
 & & \\
\dot{\vE_0} &=& -N^{-\frac{1}{2}}\nabla_s^*\diva{B}\;\; .
\end{array}
\ee

\section{Examples}
In this section we apply the general contruction, presented in the previous
sections, to two examples, which are, more or less, standard (toy) examples in
non-commutative geometry applied to elementary particle physics. In first
one the algebra $\cAs$ is a sum of two identical finite dimensional algebras
of complex matrices. This is basicly the setting of the
``Two-Point Space'' as it was presented in \cite{cobuch}. The ``Yang-Mills''
on this discrete space generates a Higgs potential and spontaneous symmetry
breaking.

In the second example the algebra of the first example is enlarged by the
algebra of smooth functions on a compact Riemannian manifold. This leads to
a gauge theory with conventional gauge bosons and Higgs bosons. The gauge
symmetry of the model is $U(n)\times U(n)$ which is broken to $U(n)$.
One might intepret this example as a model with a left-right chiral symmetry
which is broken spontaneously to a vector symmetry. However, since we do not
yet have fermions included in our construction, such an interpretation might
be a little bit artificial.

\subsection{The Two-Point Space}
We start with the discrete space and take for $\cAs$
\be
\cAs= \C^{n\times n} \oplus \C^{n\times n}
\ee
which represents the space-like part of the algebra $\cA$ in this example.
A general discussion of Connes' generalized differential algebra constructed
out of matrix-algebras can be found in \cite{KPPW}.

The complete algebra $\cA$ over space-time is then
\be
\cA = C^\infty(\R,\C^{n\times n} \oplus \C^{n\times n})\; .
\ee
The Hilbert-space $\cHs$ is
\be
\cHs = (\C^n\oplus\C^n) \otimes \C^G\otimes \C^2
\ee
where $\C^G$ denotes the ``generation-space'' with $G>1$ and the $\C^2$ factor
is needed for the construction on $\gamma^0$. The representation $\pi_s$ is
given for all $A=(A_1,A_2)\in\cAs$ as
\be
\pi_s(A)=
\left( \begin{array}{cc}
A_1 & 0 \\
    &   \\
 0  &  A_2
\end{array}\right)
\otimes 1_{\C^G}\otimes 1_{\C^2}\;\; . \label{repmm}
\ee
We take for the space-like operator $D_s$
\be
D_s=\left(\begin{array}{cc}
0   & \tilde{D}_s \\
    &             \\
\tilde{D}_s & 0
\end{array}\right) \;\; , \;\;
\tilde{D}_s=\left(\begin{array}{cc}
0   & \mu \\
    &      \\
\mu^\dagger & 0
\end{array}\right)\otimes M
\ee
where $M\in \C^{G\times G}, M^2\neq \alpha 1_{\C^{\!G}}\, , M^2\neq 0$ is a
matrix in generation space which guarantees that the representation of
two-forms on $\cHs$ is linear independent from the representation of $\cAs$.
We choose $\mu\in \C^{n\times n}$ such that
$\mu\mu^\dagger=\mu^\dagger\mu=\lambda^2 1_{\C^{\!n}}$. Thus the space-like
k-cycle $(\cHs, D_s)$ over $\cAs$ is defined and the extension to a k-cycle
$(\cH, D)$ over $\cA$ along the lines described in sect.3 is straightforward:
\be
\begin{array}{rcl}
\cH &=&L_2(\R ,(\C^n\oplus\C^n) \otimes \C^G\otimes \C^2)\;\; ,\\
 & & \\
D &=& D_t + D_s \;\; , \;\; D_t=
\left(\begin{array}{cc}
 1\partial_t & 0 \\
   & \\
0 & -1\partial_t
\end{array}\right)\;\; ,
\end{array}
\ee
where the $1$ in the definition of $D_t$ refers to the unit in
$\C^{2n}\otimes\C^G$.
The representation $\pi$ maps elements of $\cA$ onto time-dependent
blockdiagonal elements of the same form as in eq.(\ref{repmm}). The remaining
element in the general set-up which we have to specify is the $\cA$-module
$\cE$. We take the simplest choice, i.e., $\cE =\cA$.

Now we can write down the connection one form $\vA=\vA_t + \vA_s$:
\be
\vA_t = \left(\begin{array}{cc}
\vA_1 & 0 \\
    &   \\
 0  & \vA_2
\end{array}\right)\;\; ,\;\;
\vA_s= \left(\begin{array}{cc}
  0 & \phi \\
    &      \\
\phi^\dagger & 0
\end{array}\right)\;\; .
\ee
$\vA_1,\vA_2$ are anti-hermitean $n\times n$ matrices multiplied by $dt$ and
$\phi$ is a complex $n\times n$ matrix of (matrix-) form degree 1, i.e., it
is a $n\times n$ matrix multiplied by $M$.

The curvature $\vF=\vF_{st} + \vB$ of $\vA$ is given by
\be
\begin{array}{rcl}
\vF_{st} & = &\left(\begin{array}{cc}
 0 & \dot{\phi}dt + \vA_1(\mu +\phi) + (\mu + \phi)\vA_2\\
 & \\
\dot{\phi}^\dagger dt +(\mu^\dagger +\phi^\dagger)\vA_1 +
\vA_2(\mu^\dagger +\phi^\dagger)& 0
\end{array}\right)\; ,\\
 & & \\
\vB & = & \left(\begin{array}{cc}
\phi\mu^\dagger + \mu\phi^\dagger +\phi\phi^\dagger & 0 \\
 & \\
0 & \phi^\dagger\mu  +\mu^\dagger\phi +\phi^\dagger\phi
\end{array}\right)\; .
\end{array}
\ee
Since the space-like part $\cAs$ of the algebra $\cA$ is finite dimensional
the Dixmier-trace in the definition of the Lagrange function reduces to the
normal trace and hence the Lagrange function $L$ is
\be
\begin{array}{rcl}
L &=&-{1\over 4} tr (F^\dagger F)\\
 & & \\
 &=&\!\!\! {1\over 2}tr[(\dot{\phi}\gamma^0 + \vA_1(\mu+\phi)+(\mu+\phi)\vA_2)
(\dot{\phi}^\dagger \gamma^0 +(\mu^\dagger +\phi^\dagger)\vA_1 +
\vA_2(\mu^\dagger +\phi^\dagger))] - V(\phi)
\end{array}
\ee
with
\be
V(\phi)={1\over 4}tr[
\phi\mu^\dagger + \mu\phi^\dagger +\phi\phi^\dagger)
(\phi^\dagger\mu  +\mu^\dagger\phi +\phi^\dagger\phi)]\; .\label{vpot}
\ee

Now we turn to the Hamilton formalism and find for the momentum $\vE$
\be
\vE=\left(\begin{array}{cc}
 0 & -\pi^\dagger \\
  & \\
\pi & 0
\end{array}\right)
\ee
with
\be
\pi= \dot{\phi}^\dagger dt + (\mu^\dagger+\phi^\dagger)\vA_1 +
\vA_2(\mu^\dagger +\phi^\dagger)\; . \label{phimpuls}
\ee
Thus the Hamiltonian $H=H_0-G(\vA_t)$ is given by
\bes
H_0 &=& tr(\pi^\dagger\pi) + V(\phi)\\
G(\vA_t) &=& tr [\vE(D_s + \vA_s)\vA_t+ \vE\vA_t(D_s +\vA_s)]
\ees
The Gau\ss-law constraints
\be
G((\Lambda_1,\Lambda_2)dt)=0\; ,\; (\Lambda_1,\Lambda_2)=\Lambda\in\cAs,\;
\Lambda^\dagger =-\Lambda
\ee
generate the Lie-algebra of the $U(n)\times U(n)$ symmetry group.
The phase-space variables transform as follows
\be
\begin{array}{rcl}
\delta\pi &=& \Lambda_2\pi - \pi\Lambda_1\\
 & & \\
\delta\phi&=&\Lambda_1(\phi+\mu)-(\phi+\mu)\Lambda_2\; .
\end{array}\label{phitrafo}
\ee
The inhomogeneous transformation property of $\phi$ is due to the fact that
$\phi$ is part of the connection in this formalism. However, a substitution
\be
\varphi=\phi+\mu
\ee
lead to a homogeneous transformation property
\be
\delta\varphi = \Lambda_1 \varphi-\varphi\Lambda_2\;\; .\label{trvaph}
\ee
The potential $V$ reads in this new variable
\be
V(\varphi)={1\over 4}tr(\varphi\varphi^\dagger -\lambda^2)
(\varphi^\dagger\varphi-\lambda^2)\;\; .
\ee
For the time-evolution of the system we find
\be
\begin{array}{rcl}
\dot{\varphi}&=&\pi^\dagger\\
 & & \\
\dot{\pi}&=& -{1\over 4}[\varphi^\dagger(\varphi^\dagger\varphi -\lambda^2)
             + (\varphi\varphi^\dagger -\lambda^2)\varphi^\dagger]
\end{array}
\ee
We see that there are two configurations in phase-space, which are stable
under time evolution. The first one is $\pi=0, \varphi=0$, which is metastable
and $\pi=0, \varphi^\dagger\varphi=\lambda^2$ which is stable. The second
configuration is the vacuum expectation value of the Higgs-field. By
choosing for the vacuum expectation value $\varphi_0$
\be
\varphi_0=1\lambda
\ee
we infer from the transformation rule (\ref{trvaph}) of $\varphi$ that the
little group of $\varphi_0$ is the diagonal $U(n)$ subgroup of
$U(n)\times U(n)$. This shows that Yang Mills theory on discrete space
generates spontaneous symmetry breaking and thus we have translated this
appealing result of A.~Connes and J.~Lott \cite{colo} into the Hamilton
formalism.

\subsection{Yang-Mills Thoery on Space-Time $\times$ Two-Point Space}
In this second example we utilize the result of the previous example to
construct a Yang-Mills theory with spontaneously broken symmetry on a four
dimesional Minkowskian space-time. We assume that the space-time manifold
$M$ has the topology $M_3\times \R$ where $M_3$ is a compact manifold.
For this example let us take for $M_3$ the one point compactification of
$\R^3$, i.e. $M_3=S^3$. The algebra $\cA$ is of the form
\be
\cA= C^\infty(M)\otimes (\C^{n\times n} \oplus\C^{n\times n})\; .
\ee
The space-like part of the algebra is
\be
\begin{array}{rcl}
\cA_s&=& C^\infty(S^3)\otimes (\C^{n\times n} \oplus\C^{n\times n})\\
 & & \\
 &=& C^\infty(S^3)\otimes \cA_{mat}
\end{array}
\ee
For $S^3$ there is a k-cycle $(\cH_3, D_3)$ over $C^\infty(S^3)$, where $\cH_3$
denotes the square integrable spin-sections over $S^3$ and $D_3$ denotes the
Dirac-operator on $S^3$, which leads to the usual de Rham algebra. The k-cycle
$(\cH_{mat}, D_{mat})$ has been specified in the previous example (the
subscript $_{mat}$ is introduced in order to distinguish objects refering to
the discrete part of the algebra from the other objects). Usually one obtains
a k-cycle over an algebra which is a tensor product of two algebras by taking
the product k-cycle of the k-cycles over the factor algebras. However, there
is one difficulty in our case. For the definition of the operator $D$ of the
product k-cycle one needs a grading on one of the factor k-cycles. Since $S^3$
is odd-dimensional there is no such grading on the Clifford-bundle over $S^3$.
On the other hand for the Clifford-bundle over $\R\times S^3$ there is grading
given by $\gamma^5=i\gamma^0\gamma^1\gamma^2\gamma^3$. Thus we can take the
product k-cycle $(\cH, D)$ over $\cA$ with
\be
\begin{array}{rcl}
D&=&D_4\otimes 1_{mat} + \gamma^5\otimes D_{mat}\\
 & & \\
\cH&=&\cH_4 \oplus\cH_{mat}\;\; ,
\end{array}
\ee
where $D_4=\gamma^\mu\partial_\mu$ denotes the Dirac operator on
$M=\R\times S^3$ and $\cH_4$ is the space of square integrable spin-sections
over $M$. Since a Dirac operator on a manifold with topology $\R\times M_3$
can always be decomposed in a time-like part $D_t$ and a space-like part $D_3$
the space-like k-cycle $(\cHs, D_s)$ over $\cAs$ is
\be
\begin{array}{rcl}
D_s&=&D_3\otimes 1_{mat} + \gamma^5\otimes D_{mat}\\
 & & \\
\cHs&=&\cH_3 \otimes\cH_{mat}\;\; .
\end{array}
\ee
Again we choose for the $\cA$-module $\cE=\cA$.

The connection $\cA=\vA_t+\vA_s$ for this model is
\be
\vA_t = \left(\begin{array}{cc}
{\vA_t}_1 & 0 \\
    &   \\
 0  & {\vA_t}_2
\end{array}\right)\;\; ,\;\;
\vA_s= \left(\begin{array}{cc}
 {\vA_s}_1 & \phi \\
    &      \\
\phi^\dagger & {\vA_s}_2
\end{array}\right)\;\; .
\ee
$A_t$ is the same as in the previous example but on the block-diagonal of
$\vA_s$ there are now the space-like parts of the conventional gauge
connections $\vA_1$ and $\vA_2$, i.e., ${\vA_s}_1$ and ${\vA_s}_2$ are
anti-hermitean matrices multiplied with space-like one forms.

The corresponding curvature is
\be
\begin{array}{rcl}
\vF_{st} & = &\left(\begin{array}{cc}
{\vF_{st}}_1
& \dot{\phi}dt + {\vA_t}_1(\mu +\phi) + (\mu + \phi){\vA_t}_2\\
 & \\
\dot{\phi}^\dagger dt +(\mu^\dagger +\phi^\dagger){\vA_t}_1 +
{\vA_t}_2(\mu^\dagger +\phi^\dagger)& {\vF_{st}}_2
\end{array}\right)\\
 & & \\
\vB & = & \left(\begin{array}{cc}
\vB_1+\phi\mu^\dagger + \mu\phi^\dagger +\phi\phi^\dagger &
 \partial_i\phi dx^i + {\vA_s}_1(\mu +\phi) + (\mu + \phi){\vA_s}_2 \\
 & \\
\partial_i\phi^\dagger dx^i +(\mu^\dagger +\phi^\dagger){\vA_s}_1 +
{\vA_s}_2(\mu^\dagger +\phi^\dagger)
 & \phi^\dagger\mu  +\mu^\dagger\phi +\phi^\dagger\phi
\end{array}\right)
\end{array}
\ee
where $\vB_i, i=1,2$ denotes the space-like curvature of $\vA_i$,
${\nabla_s}_i$ is the corresponding covariant space-like derivative and
\be
{\vF_{st}}_i=-\partial_t{\vA_s}_idt +{\nabla_s}_i{\vA_t}_i\;\; .
\ee

Due to A.~Connes' trace theorem the Dixmier trace is in this case equivalent
to an integration over $S^3$ and hence the Lagrange function is
\be
\begin{array}{rcl}
L &=&-{1\over 4} tr\int d^3x (F^\dagger F)\\
 & & \\
 &=&{1\over 2}\int d^3x \left( -V(\phi) + tr[{\vF_{st}}_1^2 + {\vF_{st}}_2^2-
\vB_1^2 -\vB_2^2\right. \\
 & & \\
 & & +(\dot{\phi}\gamma^0 +{\vA_t}_1(\mu+\phi)+(\mu+\phi){\vA_t}_2)
(\dot{\phi}^\dagger \gamma^0 +(\mu^\dagger +\phi^\dagger){\vA_t}_1 +
{\vA_t}_2(\mu^\dagger +\phi^\dagger))\\
 & & \\
 & &\left. -(\partial_i\phi\gamma^i +{\vA_s}_1(\mu+\phi)+(\mu+\phi){\vA_s}_2)
(\partial_i\phi^\dagger \gamma^i +(\mu^\dagger +\phi^\dagger){\vA_s}_1 +
{\vA_s}_2(\mu^\dagger +\phi^\dagger))]\right)
\end{array}
\ee
with $V(\phi)$ given by eq.(\ref{vpot}).

The canonical momenta for this system are
\be
\vE=\left(\begin{array}{cc}
\vE_1 & -\pi^\dagger \\
  & \\
\pi & \vE_2
\end{array}\right)
\ee
with $\pi$ defined in eq.(\ref{phimpuls}) and
\be
\vE_i={\vF_{st}}_i\;\; ,\;\; i=1,2\;\; .
\ee
Thus we can determine the Hamiltonian $H_0 - G(\vA_t)$ to be
\be
\begin{array}{rcl}
H_0 &=& \int d^3x \left(V(\phi) +tr[ \vE_1^2 + \vE_2^2+ \pi^\dagger\pi
+\vB_1^2 +\vB_2^2 \right.\\
 & & \\
 & & \left. +(\partial_i\phi\gamma^i +{\vA_s}_1(\mu+\phi)+(\mu+\phi){\vA_s}_2)
(\partial_i\phi^\dagger \gamma^0 +(\mu^\dagger +\phi^\dagger){\vA_s}_1 +
{\vA_s}_2(\mu^\dagger +\phi^\dagger))]\right).
\end{array}
\ee
Again the Gau\ss-law can be summarized as
\be
G(\vA_t) = \int d^3x tr [\vE(D_s + \vA_s)\vA_t+ \vE\vA_t(D_s +\vA_s)]\; .
\ee
The phase-space variables transform as follows
\be
\delta\vE_i = \Lambda_i\vE_i-\vE_i\Lambda_i \; ,\; i=1,2 .
\ee
The transformation rule for the fields $\pi$ and $\phi$ are determined by
eq.(\ref{phitrafo}). By shifting $\phi$ to $\varphi=\phi+\mu$ we obtain a
field which transforms homogeneously under gauge transformations. For
$\varphi^\dagger\varphi= 1\lambda^2 $ the potential is minimized and thus
the symmetry is spontaneously broken. In the gauge
\be
\varphi = 1\lambda
\ee
we see that $\vA_+ =\vA_1 +\vA_2$ correspond to the massless modes of the
gauge fields and $\vA_-=\vA_1-\vA_2$ correspond to the massive modes.
\section{Conclusions}
We have derived the Hamilton formalism for Yang-Mills theory in non-commutative
geometry. For this purpose we exploited the special structure of
$\cA=C(I,\cAs)$ which seems to be very natural since the topology of space-time
in the conventional Hamilton formalism is $M=\R\times\Sigma$. The first step
was to show that the structure of the algebra together with an appropriately
choosen k-cycle allows to identify the time-like part of the generalized
differential algebra. Thus the notion of time obtains a well defined meaning in
this context.

The next step was to introduce the non-commutative generalization of
integration over space-like surfaces via the Dixmier trace. This opened the
possibility to apply the formalism to Minkowskian space-time by abandoning
the ellipticity of the operator $D$ of the k-cycle $(\cH,D)$ over $\cA$
but maintaining the ellipticity of the space-like part $D_s$ of $D$.
However, in this case one is restricted to the non-commutative counterpart of
integration over space-like surfaces. For the definition of Lagrange functions
and Hamilton functions integration over space-like surfaces is sufficient.
For the definition of actions one may use a hybrid formalism, i.e., one
performs integration over the (possibly non-commutative) space-like surface
via Dixmier trace and for the time variable one uses conventional integration.
The structure $C(I,\cAs)$ of the algebra ensures that this is possible.

For the definition of the Poisson bracket we had to make some additional
assumptions which we introduced at the end of sect.~5. Especially the
assumption which allowed us to define the adjoint of the operator $d$ seems
to be a brute force assumption. Although all assumptions we made are
fulfilled for the examples we presented, a finer criterion for the existence
of an adjoint of $d$ seems to be desirable.

\bye